\documentclass[apj]{emulateapj}
\usepackage{apjfonts}

\shorttitle{Cosmic Evolution of ETGs}
\shortauthors{L. Fan et al.}

\begin{document}

\title{Cosmic Evolution of Size and Velocity Dispersion\\ for Early Type Galaxies}
\author{L. Fan\altaffilmark{1,2}, A. Lapi\altaffilmark{3,1},  A. Bressan\altaffilmark{4,1}, M. Bernardi\altaffilmark{5}, G. De Zotti\altaffilmark{4,1}, and L.
Danese\altaffilmark{1}} \altaffiltext{1}{Astrophysics Sector,
SISSA, Via Bonomea 265, 34136 Trieste, Italy}
\altaffiltext{2}{Center for Astrophysics, Univ. of Science and
Technology of China, 230026 Hefei, China} \altaffiltext{3}{Dip.
Fisica, Univ. `Tor Vergata', Via Ricerca Scientifica 1, 00133
Roma, Italy} \altaffiltext{4}{INAF-Osservatorio Astronomico di
Padova, Vicolo dell'Osservatorio 5, 35122 Padova, Italy}
\altaffiltext{5}{Dept. of Physics \& Astronomy, Univ. of
Pennsylvania, 209 S. 33rd St. Philadelphia, PA 19104, USA}

\begin{abstract}
Massive (stellar mass $M_{\star}\ga 3\times 10^{10}\,
M_{\odot}$), passively evolving galaxies at redshifts $z\ga 1$
exhibit on the average physical sizes smaller by factors
$\approx 3$ than local early type galaxies (ETGs) endowed with
the same stellar mass. Small sizes are in fact expected on
theoretical grounds, if dissipative collapse occurs. Recent
results show that the size evolution at $z\la 1$ is limited to
less than $40\%$, while most of the evolution occurs at $z\ga
1$, where both compact and already extended galaxies are
observed and the scatter in size is remarkably larger than
locally. The presence at high redshift of a significant number
of ETGs with the same size as their local counterparts as well
as of ETGs with quite small size ($\la 1/10$ of the local one),
points to a timescale to reach the new, expanded equilibrium
configuration of less than the Hubble time $t_H(z)$. We
demonstrate that the projected mass of compact, high redshift
galaxies and that of local ETGs within the {\it same physical
radius}, the nominal half$-$luminosity radius of high redshift
ETGs, differ substantially, in that the high redshift ETGs are
on the average significantly denser. This result suggests that
the physical mechanism responsible for the size increase should
also remove mass from central galaxy regions ($r\la 1$ kpc). We
propose that quasar activity, which peaks at redshift $z\sim
2$, can remove large amounts of gas from central galaxy regions
on a timescale shorter than, or of order of the dynamical one,
triggering a puffing up of the stellar component at constant
stellar mass; in this case the size increase goes together with
a decrease of the central mass. The size evolution is expected
to parallel that of the quasars and the inverse hierarchy, or
downsizing, seen in the quasar evolution is mirrored in the
size evolution. Exploiting the virial theorem, we derive the
relation between the stellar velocity dispersion of ETGs and
the characteristic velocity of their hosting halos at the time
of formation and collapse. By combining this relation with the
halo formation rate at $z\ga 1$ we predict the local velocity
dispersion distribution function. On comparing it to the
observed one, we show that velocity dispersion evolution of
massive ETGs is fully compatible with the observed average
evolution in size at constant stellar mass. Less massive ETGs
(with stellar masses $M_{\star}\la 3\times 10^{10}\,
M_{\odot}$) are expected to evolve less both in size and in
velocity dispersion, because their evolution is ruled
essentially by supernova feedback, which cannot yield winds as
powerful as those triggered by quasars. The differential
evolution is expected to leave imprints in the size vs.
luminosity/mass, velocity dispersion vs. luminosity/mass,
central black hole mass vs. velocity dispersion relationships,
as observed in local ETGs.
\end{abstract}


\section{Introduction}
Most of the massive (stellar mass $M_{\star}\ga 3\times
10^{10}\, M_{\odot}$), passively evolving, galaxies at $z\ga 1$
observed with high enough angular resolution exhibit
characteristic sizes of their stellar distributions much more
compact than local early type galaxies (ETGs) of analogous
stellar mass (Ferguson et al. 2004; Trujillo et al. 2004, 2007;
Longhetti et al. 2007; Toft et al. 2007; Zirm et al. 2007; van
der Wel et al. 2008; van Dokkum et al. 2008; Cimatti et al.
2008; Buitrago et al. 2008; Damjanov et al. 2009). This very
interesting property of massive ETGs adds to others important
features: (i) luminosity, half-luminosity (or effective) radius
$R_e$ and velocity dispersion $\sigma$ of ETGs fall in a narrow
range around the so called Fundamental Plane (Djorgovski \&
Davis 1987; Dressler et al.  1987); (ii) the color-magnitude
(e.g., Visvanathan \& Sandage 1977; Sandage \& Visvanathan
1978; Bower et al. 1992a) and color-$\sigma$ (Bower et al.
1992b; Bernardi et al. 2005) relations; (iii) the increasing
$\alpha$-enhancement with increasing mass (see the discussion
by Thomas et al. 1999); (iv) the generic existence of a
supermassive black hole (BH) in their centers with mass
$M_{\bullet}\approx 10^{-3}\,M_{\star}$ (Magorrian et al. 1998;
see Ferrarese \& Ford 2005 for a review).

The first three properties imply that massive ETGs are old
systems, formed at $z_{\rm form} \ga 1.5$ on a timescale
shorter than $1$ Gyr; the environment plays a minor, but
non-negligible, role, ETGs in lower density environments being
only about $1-2$ Gyr younger (for a review see Renzini 2006).
Such properties are extremely demanding for any scenario of
galaxy formation, in particular if one sticks to the hierarchy
implied by the primordial power spectrum imprinted on dark
matter (DM) perturbations. On the other hand, the physics of
baryons (i.e., their cooling/heating mechanisms and related
feedback processes) has to play a fundamental role in galaxy
formation (e.g., Larson 1974a,b; White \& Rees 1978). The
baryon condensation in cold gas and stars within galactic DM
halos is the outcome of complex physical processes, including
shock waves, radiative and shock heating, viscosity, radiative
cooling.

In addition, the linear relationship between the central
supermassive BH and the stellar component of ETGs (point (iv)
above) can be the result of the gas removal by large
quasar-driven winds (e.g., Silk \& Rees 1998). On one side,
this hypothesis increases the complexity of the galaxy
formation process, since star formation, BH accretion, gas
inflow and outflow are interconnected and occur on quite
different space and time scales. On the other side, this
additional ingredient is very helpful. In fact, Granato et al.
(2001, 2004) show that quasar-driven winds (also named quasar
feedback) can  explain the observed $\alpha$-enhancement of
massive ETGs, the large number of submillimeter-selected
galaxies showing huge star formation rates $\dot {M}_{\star}\ga
1000\, M_{\odot}$ yr$^{-1}$ (e.g., Serjeant et al. 2008; Dye et
al. 2008) and the presence of massive, passively evolving
galaxies at $z\ga 1.5$. They also demonstrate that quasar winds
are very effective in modifying the hierarchy followed by the
assembling of DM halos, as they can account for the shorter
periods of star formation in more massive galaxies as required
by the observed galaxy stellar mass functions of ETGs, which
clearly show evidence of the so called downsizing (Cowie et al.
1996; P{\'e}rez-Gonz{\'a}lez et al. 2008; Serjeant et al.
2008). Recently, quasar feedback has been included in almost
all semianalytic models and numerical simulations of galaxy
formation, though with different recipes (see Springel et al.
2005; Croton et al. 2006; Sijacki et al. 2007; Somerville et
al. 2008; Johansson et al. 2009).

As a matter of fact, observations of $z\ga 1$ galaxies
exhibiting high star formation rates find evidence of gas in
various states, from molecular to highly ionized, with mass of
the same order of the stellar mass (Cresci et al. 2009; Tacconi
et al. 2008, 2010). If such large amounts of gas are removed
during the quasar activity, then large outflows of metal
enriched gas are expected. Such massive outflows (with rates
$\dot {M}_{\rm out}\ga 1000 M_{\odot}$ yr$^{-1}$)  have been
tentatively detected around quasars  (e.g. Simcoe et al. 2006;
Prochaska \& Hennawy 2009; L{\'{\i}}pari et al. 2009).
D'Odorico et al. (2004), studying narrow absorption line
systems associated to six quasars, have shown that these
outflows have chemical composition implying rapid enrichment on
quite short timescales (see also Fechner \& Richter 2009). The
ejection of most of the baryons initially present in
protogalactic halos is obviously necessary to explain the much
lower baryon to DM ratio in galaxies compared to the mean
cosmic value.

Fan et al. (2008) argued that rapid expulsion of large amounts
of gas by quasar winds destabilizes the galaxy structure in the
inner, baryon dominated regions, and leads to a more expanded
stellar distribution. An alternative explanation of the
increase in galaxy size calls in minor mergers on parabolic
orbits that mainly add stars in the outer parts of the galaxies
from $z\sim 2$ down to the present epoch (e.g., Nipoti et al.
2003; Hopkins et al. 2009b; Naab et al. 2009).

On the other hand, the fit to the luminosity profile of
high-$z$ galaxies may miss the outer fainter regions biasing
the size estimates (see Hopkins et al. 2010; Mancini et al.
2010); however, a detailed analysis of a galaxy at $z=1.91$ by
Szomoru et al. (2010) find no evidence of faint outer
envelopes.

In this paper we discuss critically the ideas proposed so far.
In \S~2 we give arguments leading to expect that sizes of ETG
progenitors are in fact as small as those observed; we also
discuss the data on ETG sizes as function of redshift, pointing
out the possibility that the size increase exhibits two
distinct regimes. In \S~3 we discuss the evolution of the
velocity dispersion. In \S~4 we present the relevant details on
the physical mechanisms invoked to inflate ETGs by mass loss.
In \S~5 we discuss our results, while in \S~6 we summarize our
main conclusions.

Throughout the paper we adopt the concordance cosmology (see
Komatsu et al. 2009), i.e., a flat universe with matter density
parameter $\Omega_M=0.3$ and Hubble constant $H_0= 70$ km
s$^{-1}$ Mpc$^{-1}$. Stellar masses in galaxies are evaluated
by assuming the Chabrier's (2003) initial mass function (IMF).

\section{Cosmic evolution in size of ETGs}\label{sect:sizevol}

In this section we present the recent observational evidence on
the cosmic size evolution of ETGs, and then show that small
sizes are indeed expected at high redshift if dissipationless
collapse of the baryons occurred.

\begin{figure*}
\plotone{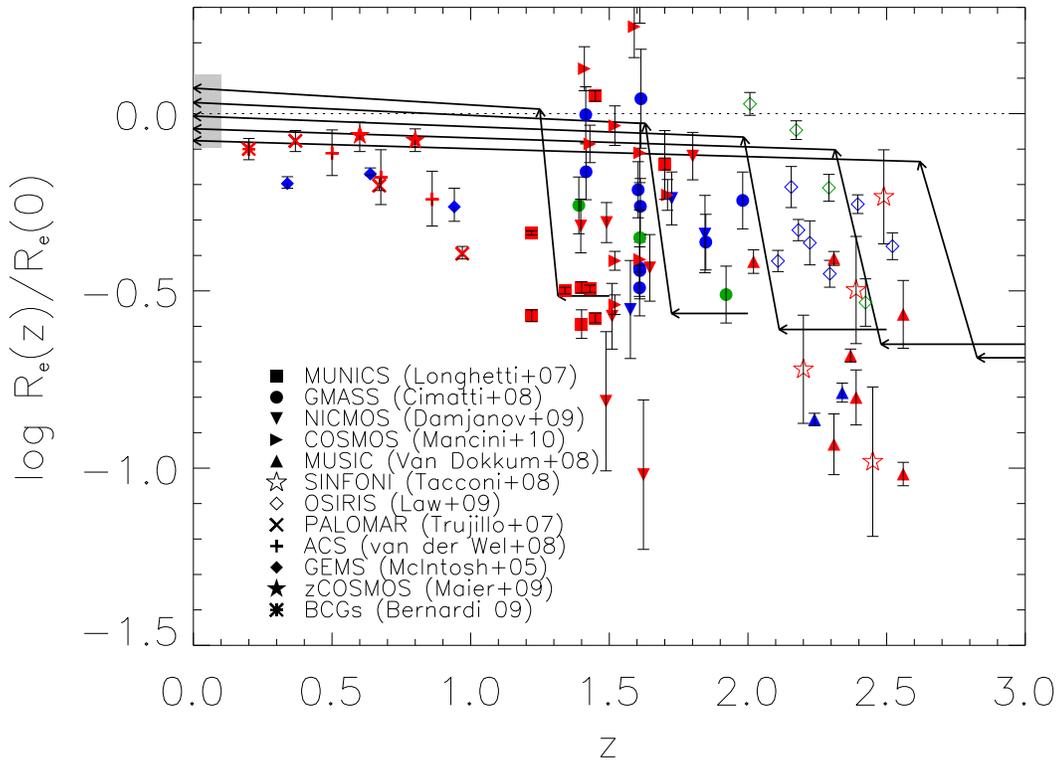}
\caption{Evolution of the effective radius with redshift. The
data points show: average sizes of $z\la 1$ passively evolving
galaxies, divided by the local sizes of galaxies of equal
stellar mass, in the samples by Trujillo et al. (2007),
McIntosh et al. (2005), van der Wel et al. (2008), Maier et al.
(2009) and Bernardi (2009), with the associated errors;
individual data and error bars for passively evolving galaxies
with spectroscopic redshifts $z\ga 1$ by Longhetti et al.
(2007), Cimatti et al. (2008), Damjanov et al. (2009), Mancini
et al. (2010), and van Dokkum et al. (2008); data and error
bars for individual star forming galaxies with spectroscopic
redshifts $z\ga 2$ by Tacconi et al. (2008) and Law et al.
(2009). The color of the data
points refers to the stellar mass of the galaxy: red is for
$M_\star\ga 10^{11}\,M_{\odot}$, blue for $3\times
10^{10}\,M_{\odot}\la M_\star\la 10^{11}\,M_{\odot}$, and green
for $M_\star\la 3\times 10^{10}\,M_{\odot}$. The shaded area
reflects the distribution of local SDSS galaxies (Hyde \&
Bernardi 2009). Thick solid lines with arrows illustrate
typical evolutionary tracks of massive galaxies according to
our reference model (with $f_{\sigma}=1.5$).}\label{fig:Rez}
\end{figure*}

\subsection{Observed size evolution of passively evolving galaxies}

A recent analysis by Maier et al. (2009) of a sample including
about 1100 galaxies with S\'{e}rsic index $n\ga 2.5$, {\it
spectroscopic} redshifts in the range $0.5\la z \la 0.9$ and
stellar masses in the range $3\times 10^{10}\,M_{\odot}\la
M_\star \la 3\times 10^{11}\, M_{\odot}$ shows that the size
evolution for galaxies at $z\sim 0.7$ is within a factor
$f_r(0.7)=R_e(0)/R_e(0.7) \la 1.25$. For galaxies at $0.7\la z
\la 0.9$ the size evolution is limited to a factor $f_r(0.9)\la
1.4$. Small size evolution ($f_r\la 1.3$) for redshifts $z \la
0.8$ was previously reported by McIntosh et al. (2005) for a
sample of 728 red galaxies with S\'{e}rsic index  $n\ga 2.5$
and stellar masses in the range $3\times 10^{9}\, M_{\odot}\la
h^2\,M_\star \la 3\times 10^{11}\, M_{\odot}$; in this case,
however, the majority of redshifts were photometric. At lower
redshift, $z\approx 0.25$, the Brightest Cluster Galaxies
(BCGs) exhibit slow evolution $f_r\la 1.3$ (Bernardi 2009).

A size evolution somewhat more pronounced (around $40\%$) than
found by Maier et al. (2009) has been claimed by Trujillo et
al. (2007) for massive galaxies $M_\star\ga 10^{11}\,
M_{\odot}$ at redshift $z\approx 0.65$. However, restricting
the analysis to galaxies with {\it spectroscopic redshifts} in
the range $0.5\la z \la 0.8$ (91 galaxies with $n\ga 2.5$ and
mean stellar mass of $1.8 \times 10^{11}\, M_{\odot}$) we find
a mean effective radius of $4.94$ kpc. The mean local effective
radius for galaxies with this stellar mass is around $6$ kpc,
implying an increase by a factor $f_r(0.65)\approx 1.2$. On the
other hand, the mean effective radius decreases to $3.8$ kpc
for galaxies with the same mean mass but with average redshift
$z\approx 0.9$; in this case the size evolution amounts to a
factor $f_r(0.9)\approx 1.6$. Similar results are found by
Ferreras et al. (2009) for a sample of 195 red galaxies
selected in the redshift range $0.4 \le z \le 1.2$. They are,
on average, more compact than local galaxies with S\'{e}rsic
index $n\ga 2.5$ by a factor of only $f_r\approx 1.4$.

A stronger evolution of $R_e$ at fixed stellar mass was
reported by van der Wel et al. (2008) for a composite sample of
50 morphologically selected ETGs in the redshift range $0.8\la
z \la 1.2$. Since we are interested on the evolution at $z\la
1$ we have confined ourselves to the 20 galaxies in a massive
clusters at $z\approx 0.83$. For these we find, on average,
$f_r(0.83)\approx 1.6$, but with a substantial mass dependence:
the most massive galaxies (dynamical mass within $R_e$ of
$M_{\rm dyn}\ga 3\times 10^{11}\,M_{\odot}$) fall quite close
to the local mass vs. $R_e$ relation, while the lower mass
galaxies tend to exhibit large size evolution.

All the results mentioned above are shown in
Fig.~\ref{fig:Rez}, where we also present a compilation of the
data at redshift $z\ga 1$. We note that, while the data points
at $z\la 1$ are averages over large samples, at higher redshift
data points refer to individual galaxies.

Assuming that the average evolution of $R_e$ can be described
by a power law of the form $R_e\propto (1+z)^{\alpha}$,
Buitrago et al. (2008) find ${\alpha}=1.48$, while van der Wel
et al. (2008) obtain a lower value $\alpha=1.20$. The latter
authors also suggest a weaker evolution, corresponding to
$\alpha=0.96$, for $z\la 1$. However, even this milder
evolution is faster than indicated by the most recent data
summarized above, and especially by the most extensive and
spectroscopically complete study of Maier et al. (2009).

\begin{figure*}
\plotone{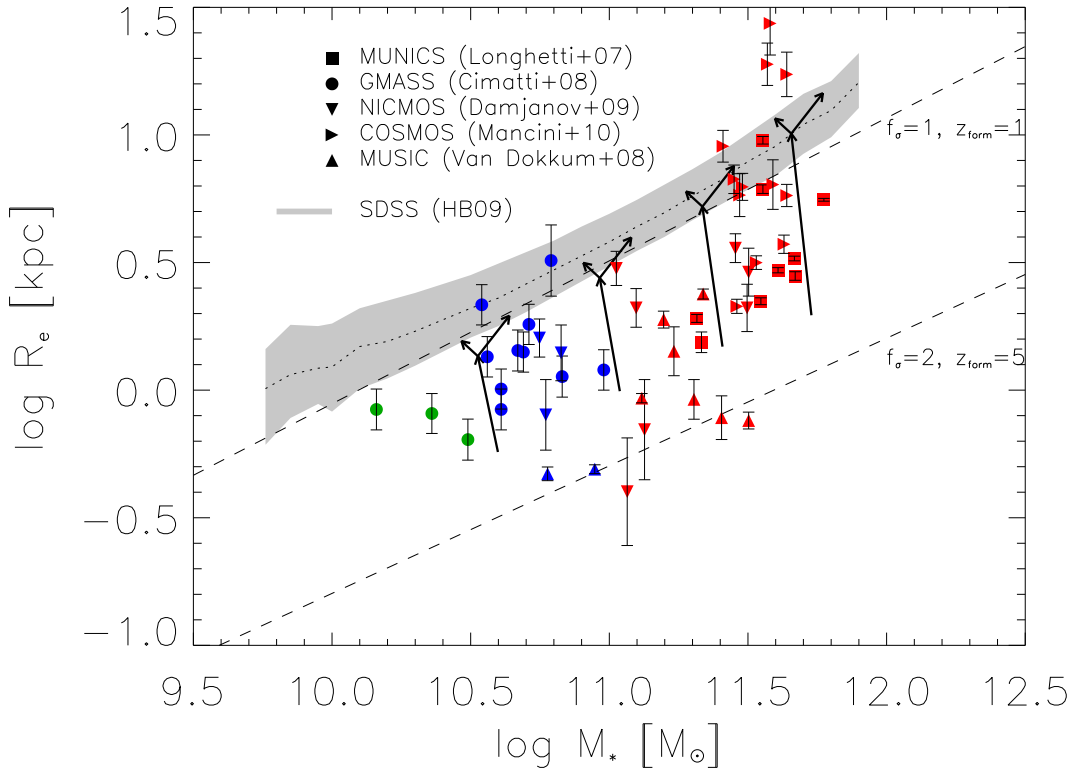}
\caption{Correlation between effective radius and stellar mass.
The observations of passively evolving galaxies with
spectroscopic redshifts $z\ga 1$ by Longhetti et al. (2007),
Cimatti et al. (2008), Damjanov et al. (2009), Mancini et al.
(2010), and van Dokkum et al. (2008) are compared with the
local correlation (Hyde \& Bernardi 2009; the dotted line
illustrates the average and the shaded area represents the
variance). Color code refers to the stellar mass, as in
previous Figure. The dashed lines illustrates the outcomes of
Eq.~(5) for extreme values of the relevant parameters
$f_{\sigma}$ and $z_{\rm form}$ (see text). Thick lines with
arrows illustrate typical evolutionary tracks of massive
galaxies according to our reference model (with
$f_{\sigma}=1.5$ and $z_{\rm form}=3$), featuring first the
abrupt size growth due to quasar feedback (almost vertical
arrows), and then the possible slow size increase due to mass
loss (arrows pointing left) or mass additions
by minor mergers (right pointing arrows.).}\label{fig:ReMstar}
\end{figure*}

A relevant feature of the data for massive galaxies at high
redshift is the quite large spread of the size, as it is
apparent from Figs.~\ref{fig:Rez} and \ref{fig:ReMstar}.
Specifically, for masses larger than $10^{11}\, M_{\odot}$ the
scatter in size of high redshift ETGs amounts to
$\sigma_{log(R_e)}\approx 0.41$, significantly wider than in
local samples, for which we have typically
$\sigma_{log(R_e)}\approx 0.14$ (cfr. Shen et al. 2003; Hyde \&
Bernardi 2009). In more detail, several high redshift galaxies
exhibit the same size as their local counterparts (see e.g.
Mancini et al. 2010; Onodera et al. 2010), while about half
ETGs exhibit $f_r\ga 3-4$, with several of them having $f_r\ga
8-10$. It is worth noticing that Maier et al. (2009) find for
the size distribution at fixed mass of their sample of ETGs at
redshift $\approx 0.7$ a statistical dispersion
$\sigma_{log(R_e)}\approx 0.16$ very close to the local one.

Provided that the presently available data constitute a
representative sample of the size of high redshift ETGs, both
the average increase of the size and the narrowing of its
distribution are to be accounted for. Only large samples of
high-$z$ ETGs will allow us to assess the interesting issue of
their size distribution. We also note that the paucity of data
at $z \ga 1$ prevents the investigation of the possible mass
dependence, a crucial aspect for any interpretation of the
phenomenon.

So far we have discussed the evolution by comparing high
redshift size determinations with the average size of local
ETGs. A bias may arise because high-$z$ samples of passively
evolving galaxies pick up objects that formed at higher
redshifts and therefore have smaller sizes. The majority of
local ETGs probably formed at $1.5 \la z_{\rm form} \la 2.5$,
but ETG progenitors already in passive evolution at $z\approx
2-2.5$ formed about $1$ Gyr earlier, i.e., at $z_{\rm form}\ga
3.5$. The latter are expected to have, on average, a factor
$\approx 1.5-2$ smaller size than local ETGs (see
Eq.~\ref{eq:retheor} below). This may explain why Valentinuzzi
et al. (2010) find that a substantial fraction (around $22\%$)
of ETGs in local galaxy clusters (overdense regions were the
galaxies typically formed earlier than in the field) are more
compact than the local average. In fact, their cluster galaxies
are on average $1.5$ Gyr older than local ETGs with `normal'
size. It is worth mentioning that several massive blue galaxies
have recently been found to exhibit compact sizes (Trujillo et
al. 2009).

\subsection{Sizes of high-redshift star forming galaxies}\label{sect:starform}

Massive starforming galaxies at high-$z$ are heavily obscured
by dust and therefore their structure cannot be investigated by
means of optical or near-IR observations; however, one can
resort to interferometric observations at millimeter and
submillimeter wavelengths. In particular CO molecular emission
has been spatially resolved for a sample of submillimeter
bright galaxies at $z\approx 2$ by Bouch\'e et al. (2007) and
Tacconi et al. (2008, 2010) with the IRAM Plateau de Bure
millimeter interferometer. The results obtained for the
galaxies with spectroscopic redshift are shown in
Fig.~\ref{fig:Rez}. For these objects the dynamical mass is a
good proxy for the mass in stars and gas.

In Fig.~\ref{fig:Rez} we also plotted the data of Law et al.
(2009) on a sample of Lyman break galaxies at redshifts $2\la z
\la 3$ and with kinematics dominated by random motions at least
in the central $2-3$ kpc. In this case, $R_e$ refers to the
$H\alpha$ or [OIII] emissions, which are sensitive to dust
extinction. The light distribution is expected to be irregular
and knotty, as in fact it is observed. Since the dust
distribution inside starforming galaxies follows the star and
gas distributions, which peak in the central regions, we expect
that the observed light profile is broadened with respect to
the true star and gas distribution for rest-frame wavelengths
shorter than a few microns (see Joung et al. 2009). Therefore
the estimated half-light radii of $H\alpha$ or [OIII] emissions
should be considered as upper limits. Nevertheless,
Fig.~\ref{fig:Rez} suggests that large starburst galaxies and
high-$z$ passively evolving galaxies, their close descendants,
exhibit the same trend of smaller size with respect to the
local ETGs.

\subsection{Expected sizes of high redshift galaxies}\label{sect:exsize}

Caon et al. (1993; see also Kormendy et al. 2009) showed that
the S\'{e}rsic (1963) function:
\begin{equation}\label{eq:sersic1}
I(r)=I(0)\,e^{-b_n\,(r/R_e)^{1/n}} ~,
\end{equation}
fits the brightness profiles of nearly all ellipticals with
remarkable precision over large dynamic ranges. Here $I(0)$ is
the central surface brightness, $R_e$ is the half-luminosity
radius, and $n$ is the S\'{e}rsic's index. The constant $b_n$
can be determined from the condition that the luminosity inside
$R_e$ is half the total luminosity $L(R_e)=L_T/2$ (see Prugniel
\& Simien 1997). The classical de Vaucouleurs (1953) profile
corresponds to $n=4$.

If light traces mass, the projected half stellar mass radius
$R_e$ is related to the gravitational radius $R_g$ by
$R_e=S_s(n)\, R_g$, and the density-weighted, $3-$dimensional
velocity dispersion $\sigma_{\star}$ is related to the observed
line-of-sight {\it central} velocity dispersion $\sigma_0$ by
$\sigma_{\star}= [3\, S_K(n)]^{1/2}\, \sigma_0$. Note that
$\sigma_0$ is usually measured within a physical size of about
$0.1\, R_e$ (e.g., J{\o}rgensen et al. 1993). At virial
equilibrium, the mass is given by:
\begin{equation}\label{eq:Mstar}
M_{\star}=S_D(n)\frac{1}{G}\, R_e\, \sigma_{0}^2~
\end{equation}
where $S_D(n) = 3\,S_K(n)/S_s(n)$. Prugniel \& Simien (1997)
have tabulated the coefficients $S_D$, $S_K$, and $S_s(n)$ for
values of the S\'{e}rsic index $n$ ranging from $1$ to $10$; in
particular, $S_s(4)\approx 0.34$, $S_K(4)\approx 0.52$, and
$S_D(4)=4.591$.

The stellar component gravitationally dominates in the inner
regions of galaxies, while the DM with its extended halo
dominates in the outer regions. To a halo with mass $M_{\rm H}$
we can associate an initial baryon mass $M_{\rm b,i}=f_b\,
M_{\rm H}$, where $f_b\approx 0.2$ is the cosmic baryon to DM
mass ratio. Weak lensing observations (Mandelbaum et al. 2006),
extended X-ray emission around ETGs (e.g., O'Sullivan \& Ponman
2004) and the comparison of the statistics of the halo mass
function with the galaxy luminosity function (Vale \& Ostriker
2004; Shankar et al. 2006; Guo \& White 2009; Moster et al.
2010) point to a present-day ratio $m=M_{\rm
H}/M_{\star}\approx 20-40$ between the total halo mass to the
total mass in stars for red massive galaxies. This result
quantifies the inefficiency of the star formation process even
in large galaxies, since $M_{\star}/M_{\rm b,i}=1/(m\,f_b)$.
The dependence of $m$ on mass and redshift predicted by our
reference model (Granato et al. 2004) is discussed in the
Appendix.

As for the inner regions, we consider the ETG progenitor
$1255-0$ at $z\approx 2.2$, for which van Dokkum et al. (2009)
and Kriek et al. (2009) find $M_{\star}\approx 2\times
10^{11}\, M_{\odot}$ within the half-light radius $R_e\approx
0.8$ kpc; local ETGs with the same mass have an average size
$R_e\approx 7$ kpc (see Shen et al. 2003; Hyde \& Bernardi
2009). If $m\approx 30$, we can associate to $1255-0$ a halo
mass $M_{\rm H}\approx 6 \times 10^{12}\,M_{\odot}$. Assuming a
NFW profile with concentration $c=4$, it is easy to see that
inside the gravitational radius $R_g\approx 3\,R_e\approx 3$
kpc, the DM fraction amounts to $f_{\rm DM}\approx 0.1$. We
notice that the DM contribution to the mass within the
half-light radius $R_e$ of local ETGs is $f_{\rm DM}\la 30\%$
(e.g., Borriello et al. 2003; Tortora et al. 2009, Cappellari
et al. 2006), and has a small effect on the stellar velocity
dispersion. This supports the notion that the dissipationless
DM cannot parallel the dissipative collapse of baryons, so that
its gravitational effects within $R_e$ can be neglected also
during the compact phase of galaxy evolution.

After the dissipative collapse of baryons inside a host DM
halo, the gravitational radius of the baryonic component, stars
plus gas with mass $M_{\star}$ and $M_{\rm gas}$ respectively,
reads
\begin{equation}\label{eq:rg}
R_g= \frac {G\, M_{\star}\left (1+f_{\rm gas}\right
)}{\sigma_{\star}^2}
\end{equation}
where $f_{\rm gas}=M_{\rm gas}/M_{\star}$ is the gas to star
mass ratio within the gravitational radius. We note that the
central gas mass includes the cold component as estimated in
the Appendix (see Eq.~[\ref{eq:Mcold}] and below for analytic
approximations).

If before collapse the baryons had the same velocity dispersion
$\sigma_{\rm b,i}$ as the DM $\sigma_{\rm DM}$, and taking into
account that $\sigma_{\rm DM}$ is approximately equal to the
halo rotational velocity $V_{\rm H}$ (see Appendix), the 3-D
stellar velocity dispersion $\sigma_{\star}$ at the end of the
collapse can be written as (Fan et al. 2008):
\begin{equation}\label{eq:sigma}
\sigma_{\star}=f_{\sigma}\, \sigma_{\rm b,i}= f_{\sigma}\,
\sigma_{\rm DM}\approx f_{\sigma}\, V_{\rm H}~.
\end{equation}
Recalling that $R_e=S_s(n)\, R_g$, with $R_g$ given by
Eq.~(\ref{eq:rg}), and that $V_{\rm H}^2=G\,M_{\rm H}/R_{\rm
H}$ we then obtain $R_e$ in terms of the halo radius $R_{\rm
H}$ and mass $M_{\rm H}$:
\begin{eqnarray}\label{eq:retheor}
R_e &\approx& \frac{S_s(n)}{f_{\sigma}^2}\,
\frac{\left(M_{\star}+M_{\rm gas}\right)}{M_{\rm H}}\, R_{\rm
H}\approx \\
\nonumber&\approx& 0.9\, \frac{S_s(n)}{0.34}\,\frac{25}{m}\,\left(\frac
{1.5}{f_{\sigma}}\right)^2\,\left(\frac{M_{\rm H}}{10^{12}\,
M_{\odot}}\right)^{1/3}\,\left(\frac{4}{1+z_{\rm
form}}\right)~~~\mathrm{kpc},
\end{eqnarray}
where $z_{\rm form}$ is the redshift when the collapse begins,
and we have set $f_{\rm gas}=1$. This equation shows that the
baryon collapse naturally leads to kpc or sub-kpc effective
radii and to stellar velocity dispersions higher than halo
rotational velocities ($f_{\sigma}>1$). Both these properties
differ from those observed for local massive galaxies, implying
that other ingredients have come into play.

The explicit redshift dependence of $R_e$ in
Eq.~(\ref{eq:retheor}) comes from the halo radius, which scales
as $(1+z_{\rm form})^{-1}$. In addition, the ratio $m$, which
measures the star formation inefficiency and is determined by
the physics of baryons, scales as $(1+z_{\rm form})^{-0.25}$
(see Appendix). As a result the effective radius scales like
$R_e\propto (1+z_{\rm form})^{-0.75}$. The values of $m$ and of
$f_{\sigma}$ depend on how and when the star formation and gas
heating processes can halt the collapse. The latter must
proceed at least until the mass inside the $R_e$ is dominated
by stars, as observed in local ETGs.

In Fig.~\ref{fig:ReMstar} we compare the observed distribution
of local and high-$z$ galaxies in the $R_e$ vs. $M_\star$ plane
with expectations from Eq.~(\ref{eq:retheor}). A robust upper
limit to the high-$z$ correlation (upper dashed line in
Fig.~\ref{fig:ReMstar}) is obtained setting $f_{\sigma}=1$
(baryon collapse with no increase of the stellar velocity
dispersion) and $z_{\rm form}=1$, corresponding to a look-back
time of about $7-8$ Gyr, a lower limit to the mass-weighted age
of local massive ETGs (Gallazzi et al. 2006; Valentinuzzi et
al. 2010). The corresponding line falls just at the lower
boundary of the distribution of local ETGs, but at the upper
boundary of the distribution of high-$z$ passively evolving
massive galaxies. The median size of the latter is a factor
around $4$ lower than that of local galaxies with the same
stellar mass. This argument is not in contrast with the
existence, recently reported by Valentinuzzi et al. (2010), of
local compact ETGs, which can represent the evolution of the
oldest, most compact progenitors.

The lower bound to the high-$z$ correlation is less well
defined; in Fig.~\ref{fig:ReMstar} the lower dashed line
corresponds to $z_{\rm form}\sim 5$ and to $f_{\sigma} = 2$. In
Eq.~(\ref{eq:retheor}), we have adopted as our reference values
$f_{\sigma}=1.5$ and $z_{\rm form}=3$.

\section{Evolution of the ETG velocity dispersions}\label{sect:vdisp}

Applying the virial theorem to the galaxies before and after
their growth in size, we have that the final line of sight
central stellar velocity dispersion $\sigma_{\rm 0,f}$ is
related to the initial one $\sigma_{\rm 0,i}$ by:
\begin{equation}
\sigma_{0,f}^2 =
\sigma_{0,i}^2\,\frac{S_D(n_i)}{S_D(n_f)}\,
\frac{M_{f}}{M_{i}}\,\frac{R_{e,i}}{R_{e,f}} = \frac
{f_{\sigma}^2\, V^2_{\rm H}(z_{\rm form})}{3\,
S_K(n_i)}\,\frac{S_D(n_i)}{S_D(n_f)}\,\frac{M_f}{M_i}\,\frac
{R_{e,i}}{R_{e,f}}~; \label{eq:sigmaVH}
\end{equation}
here the indices $i$ and $f$ label quantities in the initial
and final configuration, and $S_D(n)$ is the structure factor
defined by Prugniel \& Simien (1997; see Eq.~[\ref{eq:Mstar}]),
$n$ being the S\'{e}rsic index. In the last expression we have
used Eq.~(\ref{eq:sigma}) and the relation $\sigma_\star^2 =
3\,S_K(n_i)\,\sigma_{0,i}^2$. In the case of an homologous
growth of the galaxy size, the velocity dispersion scales as
$(M/r)^{1/2}$, so that it remains constant if both the mass and
the size increase by the same factor and decreases as
$r^{-1/2}$ if the growth occurs at constant mass. However, the
size growth is not necessarily homologous. All the mechanisms
so far proposed predict an increase of the S\'{e}rsic index
with increasing size. This effect together with a possible
increase in mass within the limits imposed by the mass function
evolution, tend to soften the decrease of the velocity
dispersion. A further attenuation of the evolution is expected
because of dynamical friction with the DM component.

The above equation shows that the size evolution of ETGs is
paralleled by velocity dispersion evolution and that the
present-day velocity dispersion keeps track of the potential
well of the host halo when the galaxy forms. This is expected,
since in a galaxy halo the gas is channeled toward the central
regions during the fast accretion phase under the effect of the
DM potential well. The duration of the star formation process
depends on halo mass, feedbacks and redshift at which the fast
accretion phase occurs. The velocity dispersion of the
collapsed galaxy is not affected by the minor fraction of DM
added subsequently to the external regions of a halo during the
slow accretion phase (see Lapi \& Cavaliere 2009).

Observations of the kinematics of passively evolving ETGs at
$z\ga 1$ are quite difficult. Nonetheless for two galaxies
reliable estimates of the velocity dispersion have been
obtained (Cappellari et al. 2009; van Dokkum et al. 2009). For
other objects only average estimates have been inferred from
stacked spectra (Cenarro \& Trujillo 2009; Cappellari et al.
2009). In all cases the main conclusion is that stellar masses
derived from spectrophotometry are in good agreement with
virial masses or with masses derived from dynamical models, if
one adopts an IMF flattening below $1\, M_{\odot}$, such as
those proposed by Kroupa (2001) or by Chabrier (2003). This
finding is also confirmed at intermediate redshifts $0.4\la
z\la 0.9$ by van der Wel et al. (2008).

One of the two high redshift ETGs with a good determination of
the velocity dispersion, GMASS 2470 (Cappellari et al. 2009),
falls in the $\sigma$ vs. $R_e$ plane quite close to the area
covered by local ETGs. On the other hand, the best fit value of
the stellar velocity dispersion, $\sigma_0=510^{+165}_{-95}$,
for the galaxy 1255-0 at $z\approx 2.2$ (van Dokkum et al.
2009) exceeds the measured values for even the most massive
local galaxies (Bernardi et al. 2008). Although we cannot do
statistics with a single case, its existence lends support to
the possibility of a significant evolution of the galaxy
velocity dispersion.

Cappellari et al. (2009; see also Bernardi 2009), based on
stacked spectra of $13$ galaxies at $1.4\la z\la 2.0$ (cf.
their Table~1), find a mild evolution of the velocity
dispersion, that decrease from about $202$ km s$^{-1}$ at $z
\approx 1.6$ down to about $160$ km s$^{-1}$ at $z\approx 0$
for $M_\star \approx 7\times 10^{10}\,M_\odot$, and find an
increase of the source size by a factor around $3.5$. This
evolution can be understood if the S\'{e}rsic's index $n$
increases from an initial value $n_i$ to $n_f=n_i+2$ and the
mass increases by $30\%$; in that case the velocity dispersion
decreases by a factor $1.35$, or less if dynamical friction
with DM has a role. Note that if, as shown in
\S~\ref{sect:exsize}, the DM is dynamically irrelevant in the
inner regions of galaxies, and mergers accrete matter in the
outer regions, the mass within $R_e$ does not change and the
same velocity dispersion evolution applies also to the minor
merger scenario.

A quite interesting upper limit to the velocity dispersion,
$\sigma_{\star}\la 326$ km s$^{-1}$, for a massive
$M_{\star}\approx 3-4 \times 10^{11}\,M_{\odot}$ at redshift
$z\approx 1.82$ has been found by Onodera et al. (2010). The
same authors also find that the size of this galaxy is as
expected for a local galaxy with the same mass. The velocity
dispersion and the size yield a virial mass upper limit
$M_{H}\la 7\times 10^{11}\, M_{\odot}$, quite close to the
stellar mass. This galaxy has the same structural properties of
a local ETG. Recalling that a significant fraction of $z \ga
1.5$ galaxies already exhibit a size close to the size of their
local counterparts, this galaxy appears as a well studied case
of an already evolved galaxy, suggesting that the timescale for
the size evolution is shorter than the Hubble time at those
redshifts $\Delta T_{\rm size}< t_H(z)$.

\section{Physical mechanisms for size evolution}

Both theory and observations suggest that at least $60\%$ of
ETGs evolve in size by at least a factor of $2-4$. So far, two
main mechanisms have been proposed to accomplish such
evolution.  One possibility is that the expansion is driven by
the expulsion of a substantial fraction of the initial baryons,
still in gaseous form, by quasar activity (Fan et al. 2008) or
by an expulsion of gas associated to stellar evolution (e.g.,
Damjanov et al. 2009). The two mechanisms differ in the
expulsion timescale, which is shorter than the dynamical time
if it is triggered by quasar activity and longer in the case of
ejection associated to stellar evolution (with `standard'
IMFs).

Alternatively, the increase in size could be due to minor
mergers on parabolic orbits that add stars in the outer parts
of the galaxies along the cosmic time from $z\approx 1-2$ to
the present epoch (see Maller et al. 2006; Naab et al. 2009;
Hopkins et al. 2009b; van der Wel et al. 2009). Major mergers
(i.e., mergers of galaxies with similar mass) can also increase
the galaxy size in a way almost directly proportional to the
mass increase and they were also considered (e.g.,
Boylan-Kolchin et al. 2006; Naab et al. 2007) but the required
space densities of progenitors were found to be incompatible
with the present-day galaxy mass function (Bezanson et al.
2009; Toft et al. 2009) as well as with the dearth of compact,
massive galaxies in the local universe (Trujillo et al. 2009).

A third possibility is that the increase is illusory, because
the low-surface brightness in the outer regions of high-$z$
galaxies may be missed and the effective radii are
correspondingly underestimated (Mancini et al. 2010; Hopkins et
al. 2010) or because a gradient in the $M/L$ ratio (lower in
the bluer central regions) can make the half-light radius in
the optical smaller than the half-mass radius (Tacconi et al.
2008); however, Szomoru et al. (2010) find no evidence of outer
faint envelopes in a well-studied galaxy at $z\approx 1.9$.

\subsection{Gas expulsion}\label{sect:exp}

In the case of gas expulsion the final size depends on the
timescale of the ejection itself. If the ejection occurs on a
timescale shorter than the dynamical timescale of the system
$\tau_{\rm ej}<\tau_{\rm dyn}$, immediately after the ejection
the size and velocity dispersion are unchanged but the total
energy is larger because the mass has decreased. The system
then expands and evolves towards a new equilibrium
configuration. In the case of homologous expansion the final
size $R_f$ is related to the initial one $R_i$ by (Biermann \&
Shapiro 1979; Hills 1980):
\begin{equation}\label{eq:BS}
\frac {R_i}{R_f}=1-\frac {M_{\rm ej}}{M_{f}}
\end{equation}
where $M_{\rm ej}$ is the ejected mass and $M_{f}$ is the final
mass.

This simple result has been confirmed by numerical simulations
of star clusters (e.g., Geyer \& Burkert 2001; Boily \& Kroupa
2003). In particular, the simulations by Goodwin \& Bastian
(2006) and by Baumgardt \& Kroupa (2007) show that the
expansion of the half-mass radius occurs in about $20$
dynamical times and the new final equilibrium is attained
within $40$ dynamical times. We note that the case of galaxies
differs from that of the star clusters owing to the presence of
the DM halo. In ETGs the DM halo exerts its gravitational
influence outside the central region dominated by stars and
prevents the galaxy disruption when $M_{\rm ej}$ approaches or
exceeds $M_{f}$; the DM potential can also influence the time
taken by the stars to reach the new equilibrium.

When the mass loss occurs on a timescale longer than the
dynamical time the system expands through the adiabatic
invariants of the stellar orbits and one gets
\begin{equation}\label{eq:adexp}
\frac {R_f}{R_i}=1+\frac {M_{\rm ej}}{M_{f}}~.
\end{equation}
Comparison of the two above equations show that the fast
expulsion is more effective in increasing the size.

The dynamical time of the stellar component is
\begin{equation}
\tau_{\rm dyn}=\pi\, \left(\frac {R_e^3}{2\, G\,
M_{\star}}\right)^{1/2}\approx 3\times 10^{6}\, \left(\frac
{R_e}{1~\rm kpc}\right)^{3/2} \left(\frac{10^{11}\,
M_{\odot}}{M_{\star}}\right)^{1/2}~\mathrm{yr},
\end{equation}
about $30-50$ times shorter than the typical dynamical
timescale in local massive ETGs and not much longer than the
dynamical timescale usually associated to star clusters. In the
case of mass loss due to stellar feedback (Hills 1980;
Richstone \& Potter 1982) $\tau_{\rm ej}\gg \tau_{\rm dyn}$ for
any reasonable choice of the IMF. For instance, if a Chabrier
(2003) or Kroupa (2001) IMF is adopted, after an initial burst
about half of the mass of formed stars returns to the gaseous
phase over a timescale of $1$ Gyr. If this gas is removed from
galaxies, the size may grow by a factor of about $2$. The
higher the proportion of massive stars, the larger the effect
on the size, and the shorter the timescale for the size
expansion (see Damjanov et al. 2009).

In the case of quasar winds the typical timescale for gas
ejection can be estimated as
\begin{eqnarray}
\nonumber \tau_{\rm ej} = \frac {M_{\rm gas}}{\dot {M}_{\rm wind}} &\approx&
5\times 10^6\, \left(\frac{m}{25}\right)^{2/3}\,\left
(\frac{M_\star}{10^{11}\, M_{\odot}}\right)^{5/3}\times \\
&\times& \left(\frac{M_{\bullet}}{2\times 10^{8}\, M_{\odot}}\right
)^{-3/2}\,\left(\frac{1+z}{4}\right)~\mathrm{yr},
\end{eqnarray}
where $\dot{M}_{\rm wind}$ is given by Eq.~(A12), and we have
assumed $M_{\rm gas}\approx M_\star$. An alternative definition
of $\tau_{\rm ej}$ is
\begin{equation}
\tau_{ej}=\frac{R_e}{V_e}\approx 10^6\, \left(\frac{R_e}{1\,\mathrm{
kpc}}\right)^{1/2}\, \left(\frac{M_\star}{10^{11}\, M_{\odot}}
\right)^{-1/2}\, \mathrm{yr}~,
\end{equation}
where $V_e^2=2\,G\,M_\star/R_e$ is the escape velocity from the
radius $R_e$. With both definitions the ejection timescale is
of the order of the dynamical timescale.

It is apparent that numerical simulations are badly needed to
investigate the detailed effect of quasar winds on the size and
on the timescale $\Delta T_{\rm size}$ to reach the new
equilibrium; such kind of simulations are underway (L. Ciotti
and F. Shankar 2009, private communication).

\subsection{Minor mergers}\label{sect:minmer}

In the case of minor mergers on parabolic orbits the initial
potential energy of the accreting mass is neglected in the
computation. Following Naab et al. (2009) we assume that random
motions are dominant in high-$z$ ETG precursors and set
$\eta=M_a/M_i$ and $\epsilon = \sigma^2_a/\sigma^2_i$, the $i$
and $a$ indices referring to initial and accreted material. The
mass after merging is therefore $M_f=M_i\,(1+\eta)$. If $r
\propto M_{\star}^{\alpha}$, the virial theorem gives
$\epsilon=\eta^{1-\alpha}$. Local ETGs have $\alpha \approx
0.56$ (Shen et al. 2003) or even larger in the case of BCGs
(Hyde \& Bernardi 2009); in addition, a value $\alpha \approx
0.5$ is implied by the Faber-Jackson (1976) relationship.

From the virial theorem and the energy conservation equation it
is easily found that the fractional variations of the
gravitational radius and of the velocity dispersion between the
configurations before ($i$) and after ($f$) merging are:
\begin{eqnarray}\label{eq:rexp}
\nonumber\frac {R_{g,f}}{R_{g,i}}=\frac {(1+\eta )^2}{(1+ \eta^{2-\alpha})}~,\\
\\
\nonumber\frac {\sigma_{f}^2}{\sigma_{i}^2}=\frac {(1+
\eta^{2-\alpha})} {(1+\eta )}~.
\end{eqnarray}
Boylan-Kolchin et al. (2008) showed that minor mergers can be
effective only if $\eta \ga 0.1$, lower mass ratios requiring
too long timescales.

Recent numerical simulations by Naab et al. (2009) agree with
these results. Their simulated galaxy, with a mass in stars
$M_{\star}\approx 8 \times 10^{10}\, M_{\odot}$ and half-mass
radius $R_e\approx 1$ kpc at $z \approx 2$, by $z=0$ has
doubled its stellar mass through minor mergers, reaching
$M_{\star}\approx 1.5 \times 10^{11}\,M_{\odot}$, while the
half-mass radius has increased by a factor $2.7$. The
simulations also suggest that most of the increase, a factor of
about $1.8$, occurs at $z\la 1$, i.e, on a cosmological
timescale. This is accompanied by a moderate decrease, $\la
20\%$, of the central velocity dispersion between $z\approx 3$
and $z\approx 0$ and by a decrease of the central density of
stellar distribution with time, due to dynamical friction,
despite of the total mass increase. However, these simulations
yield a present-day half-mass radius a factor of $2$ smaller
than expected on the basis of the $M_{\star}$ vs. $R_e$
relationship of Shen et al. (2003; see also
Fig.~\ref{fig:ReMstar}).

We notice that the size evolution in the merging case occurs on
timescale which is comparable with the present Hubble time with
size scaling $\propto (1+z)^{\beta}$; in the simulations of
Naab et al. (2009) $\beta \approx 1$ holds, and similarly in
the findings of van Dokkum et al. (2010) $\beta \approx 1.27$.

\section{Discussion}

The observational data and the theoretical arguments summarized
in the previous sections allow us to test and constrain the
different models for size evolution. Since a size increase by
minor dry mergers implies an increase in mass, we start by
discussing limits on the latter.

\subsection{The mass evolution of ETGs}\label{sect:masslim}

Spectral properties of local ETGs with stellar masses
$M_{\star}\ga 3\times 10^{10}\,M_{\odot}$ indicate that their
light-weighted age exceeds $8-9$ Gyr, independently of the
environment (see Renzini 2006 for a review and Gallazzi et al.
2006 for an extensive statistical study). Since light-weighted
ages are lower limits to mass-weighted ages (e.g., Valentinuzzi
et al. 2010), it is generally agreed that most of the stars of
massive ETGs formed at $z_{\rm form}\ga 1.5-2$. An upper limit
$\la 25\%$ to the fraction of stars formed in ETGs in the last
$\approx 8$ Gyr, and as a consequence to the fraction of gas
accreted at intermediate redshift $z\la 1$, has been derived
from studies of narrow band indices of local field ETGs (e.g.,
Annibali et al. 2007). Moreover, all massive galaxies that
formed and gathered the bulk of their stars at $z\ga 1$ are
presently ETGs or massive bulges of Sa galaxies, since there
are no late-type, disc-dominated galaxies endowed with so large
masses of old stellar populations.

Thus by comparing the stellar mass function of local ETGs to
the mass function of all galaxies at $z\la 1.5$, we can derive
information on the mass evolution.

Bernardi et al. (2010) studied in detail about $2000$
morphologically-classified local galaxies extracted from the
SDSS sample (see Fukugita et al. 2007). They showed that the
concentration index $C_r$ can be used to discriminate among
galaxy types. The criterion $C_r>2.86$ includes almost all
ellipticals, about 80$\%$ of S$0$ galaxies and 40$\%$ of Sa
galaxies, likely those with a larger disk component of younger
and bluer stars. Correspondingly, while the fraction of E and
S0 massive galaxies ($M_{\rm dyn}\ga 10^{11}\, M_{\odot}$)
older than $8$ Gyr is quite large, the fraction of massive and
old Sa galaxies is less than $50\%$ (cf. Fig.~23 of Bernardi et
al. 2010).

\begin{figure*}
\plotone{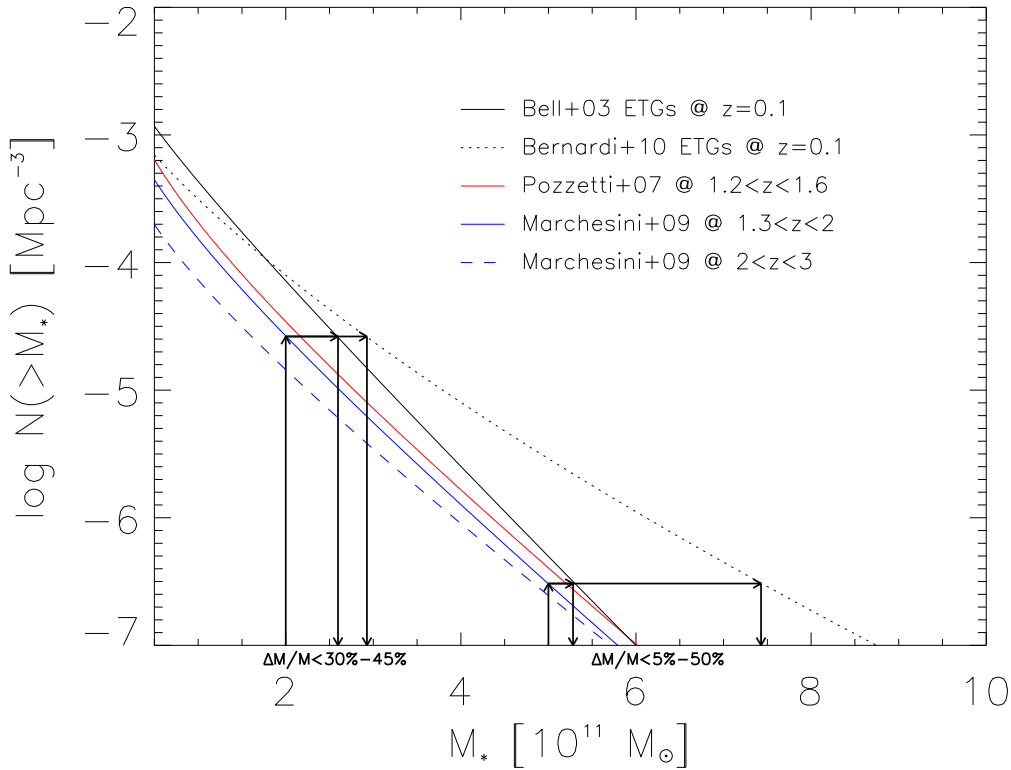}
\caption{Cumulative stellar mass function. The different lines
illustrate Schechter fits to the average stellar mass function
at different redshifts as estimated by Bell et al. (2003),
Bernardi et al. (2009), Pozzetti et al. (2007), and Marchesini
et al. (2009). All estimates have been scaled to Chabrier's
(2003) IMF. Thick solid lines with arrows highlight the mass
evolution from $z\approx 1.5$ to the present allowed by the
observed mass functions, starting from $M_{\star}\approx 2$ and
$5\times 10^{11}\, M_{\odot}$.}\label{fig:smf}
\end{figure*}

Therefore, in order to consistently compare the high redshift
mass function with the local one, in Fig.~3 we report the
cumulative mass function for the full Bernardi et al. (2010)
sample with concentration index $C_r>2.86$. We also plot the
mass function by Cole et al. (2001; similar results were
obtained by Bell et al. 2003), that has been computed with
criteria that tend to exclude late type galaxies. These local
mass functions are compared with estimates by Pozzetti et al.
(2007) and Marchesini et al. (2009), which refer to redshift
$z\approx 1.4$ and $z\approx 1.6$, respectively. All mass
functions have been rescaled to Chabrier's (2003) IMF.

When comparing the local to the high redshift ETG mass
function, the first issue is how to make a complete census of
high redshift ETG progenitors. Williams et al. (2009) found
that in a deep sample (magnitudes $K_{\rm AB} < 22.4$) the most
luminous objects at $z\sim 1-2$ are divided roughly equally
between starforming and quiescent galaxies. A significant
fraction of galaxies at $z\ga 1.5$, forming stars at rates of
hundreds to thousands solar masses per year as revealed by
far-IR or submillimeter surveys, are easily missed even by deep
$K-$band surveys because of their strong dust obscuration
(e.g., Dye et al. 2008). An extreme example is GN10, a galaxy
at $z\approx 4$ that exhibits a star formation rate around
$1000\, M_{\odot}$ yr$^{-1}$, a stellar mass around $10^{11}\,
M_{\odot}$ and a dust extinction $A_V\sim 5-7.5$ mag (see Daddi
et al. 2009; Wang et al. 2009); this went undetected by
ultradeep $K_s-$band exposures, yielding a $1-\sigma$ upper
limit of 23 nJy (Wang et al. 2009) that corresponds to $K_{\rm
AB} \ga 28$. Since all galaxies at redshift $z\ga 1.5$ have to
be included in the budget, mass functions of high redshift
massive galaxies based on optical or near-IR selected samples
should be regarded as lower limits to the high redshift
counterparts of local ETGs (see Silva et al. 2005 for a more
detailed discussion). As a consequence, only upper limits to
evolution in mass allowed by the data obtains by assuming that
the galaxy number density keeps constant.

In addition, it is apparent from Fig.~1 of van Dokkum et al.
(2010) that the upper limit to mass evolution slightly depends
on mass or correspondingly on the reference number density. van
Dokkum et al. (2010) adopt as a reference number density
$2\times 10^{-4}$ Mpc$^{-3}$ dex$^{-1}$ and find that galaxies
with this number density at $z\approx 1.6$ are endowed with
$M_{\star}\approx 1.7 \times 10^{11}\,M_{\odot}$, while at
$z\approx 0.1$ the same number density pertains to galaxies
with $M_{\star}\approx 2.8 \times 10^{11}\,M_{\odot}$ (the
adopted local number density is that of Cole et al. 2001); the
ensuing upper limit to mass evolution is $\la 70\%$. Applying
the same argument to galaxies with number density $2\times
10^{-5}$ Mpc$^{-3}$ dex$^{-1}$ yields an upper limit to mass
evolution $\la 40\%$. If the local number density of Bernardi
et al. (2010) is adopted, the upper limit to mass evolution is
$\la 50\%$ since $z\approx 1.6$ with practically no dependence
on mass, as shown in Fig.~\ref{fig:smf}.

These upper limits are compatible with evidences that most, if
not all, massive ETGs are already in place at redshift
$z\approx 1$ (see Drory et al. 2005; P{\'e}rez-Gonz{\'a}lez et
al. 2008; Fontana et al. 2006; Cirasuolo et al. 2010; Kajisawa
et al. 2009) and that only a fraction $\la 30\%$  of their
stellar mass can be added at later times. Collins et al. (2009)
have estimated the masses of the Brightest Cluster Galaxies
(BCGs) in $5$ of the most distant X-ray-emitting galaxy
clusters at redshifts $z\sim 1.2-1.5$, finding that they are
perfectly compatible with the local average mass of BCGs. If
the two galaxies, which have companions, incorporated them,
their mass would increase in one case by about $20\%$ and in
the other by $40\%$.

The results of numerical simulations on DM halos are compatible
with such a mass increase. More in detail, Boylan-Kolchin et
al. (2008) showed that only merging of satellites with mass
ratio $\eta \ga 0.1$ can efficiently increase the mass of their
host galaxies. Also the merging rate for massive galaxies
inferred from numerical simulations by Stewart et al. (2008)
confirms that most of the mass is added by merging of
satellites with mass ratio $\eta\ga 0.1$. We stress, however,
that these simulations refers to the DM halos and its
translation to stellar component of merging halos is not
trivial.

To sum up, the data allow, at most, for a mass increase by a
factor of $\approx 2$ since $z\approx 2$ and by a factor of
$\approx 1.5-1.7$ since $z\approx 1.5$. We notice also that if
the growth occurs via minor dry mergers, with no evolution of
the galaxy number density, practically all massive galaxies
gradually increased their mass throughout their entire
lifetime, from the formation redshift to $z=0$ [see, e.g., the
simulations by Naab et al. (2009) and Stewart et al. (2008)].
But then also the galaxy sizes should increase gradually over
the full galaxy lifetime, and this can be hardly reconciled
with the much larger scatter in size observed for ETG
progenitors at $z\ga 1.5$ compared to that at lower redshifts.

\subsection{Size evolution}\label{sect:sizev}

The comparison of available data on $z\ga 1$ ETGs with the
local size distribution clearly points toward a mean size
increase by about a factor of $3$ in order to bring the average
$M_{\star}$ vs. $R_e$ relationship of high redshift ETGs to the
local average (cf. Fig.~\ref{fig:Rez}), though we caution that
larger samples of high redshift ETGs are needed. We stress that
the observed small sizes at high-$z$ are indeed expected (cf.
Eq.~[\ref{eq:retheor}]) if ETG progenitors formed most of their
stars in a rapid, dissipative collapse.

As a matter of fact, we expect that high redshift passively
evolving ETGs formed at redshift $z_{\rm form}\ga 4$ larger
than the formation redshift $z_{\rm form}\approx 1.5-2$ of most
local ETGs. From the number density of halos with $M_{\rm H}\ga
10^{12}\,M_{\odot}$ as a function of redshift, we estimate that
massive galaxies (with $M_{\star}\ga 10^{11}\, M_{\odot}$)
formed at redshift $z_{\rm form}\ga 4$ are only $10\%$ of those
formed at $z_{\rm form}\approx 2$. Therefore, since $R_e\propto
(1+z_{\rm form})^{-0.75}$ (see \S~\ref{sect:exsize}), the local
counterparts of high-$z$ ETGs, $10\%$ of the total number of
ETGs, are expected to exhibit a half-light radius smaller than
the average by a factor around $1.4$.

Taking into account this bias, data in Fig.~\ref{fig:Rez} and
in Fig.~\ref{fig:ReMstar} show that a significant fraction of
local ETG precursors already at $z\ga 1.5$ exhibit the same
size as their local counterparts of the same mass. On the other
hand, there are also ETG progenitors much more compact than
their local counterparts, with sizes smaller by a factor $\la
1/6$. As a matter of fact, the dispersion in size at high
redshift is larger than in the local samples of ETGs. These
properties of the size distribution can be accounted for by a
model yielding evolution in size by large factors ($\ga 5$) on
timescales shorter than the Hubble time $t_H$ at $z\ga 1.5$.
Ejection of large amounts of gas by quasar feedback can
reproduce the observed phenomenology. From Eq.~(7) it is
apparent that large size expansions are possible, even though
gravity of DM halos will constrain them.  Also from Eq.~(9) it
is apparent that timescales from a few to several $10^8$ yr can
be required for the expansion. In this context the scatter of
sizes mirrors the spread of  formation time and the spread in
the expansion phase, as illustrated by Fig.~\ref{fig:Rez} and
Fig.~\ref{fig:ReMstar}. In particular, in Fig.~\ref{fig:Rez}
lines with an arrow represent the size evolution predicted by
our reference model. The lower horizontal lines represent the
time (translated to $\Delta z$) spent by ETG progenitors in
their dusty phase with quite large star formation rate; submm
surveys are quite efficient in selecting this phase. Then the
(almost vertical) lines represent the epoch of the large size
increase due to the gas outflow triggered by the quasar
activity; this phase begins with the quasar appearance and
lasts $\Delta T_{\rm size}$ (here $\Delta T_{\rm size}\approx
2\times 10^8$ yr has been assumed). Then a longer phase lasting
for about the present Hubble time follows, during which the
size can increase by a smaller factor because of mass loss due
to galactic winds and/or minor dry mergers.

We note that quasar feedback has not been introduced
specifically to solve the size problem, but first by Silk and
Rees (1998) to predict the correlation in ETGs between galaxy
velocity dispersion and the present-day BH mass. Soon after,
Granato et al. (2001, 2004) have shown that the gas removal by
quasar activity is also needed in order to stop the star
formation, preventing formation of exceedingly massive
galaxies, too blue and with no enhancement of $\alpha$-elements
(cf. \S~\ref{sect:masslim}). Sterilization of star formation by
quasar feedback implies that in a quite short timescale, an
enormous mass of gas is evacuated from the central galaxy
regions and possibly from the entire halo and subhalos. The gas
evacuated from the central regions can be of the same order of
the mass in stars, so about $10-20\%$ of the total baryons in
the galactic halo. Observations of  high redshift star forming
galaxies do find evidence of large fractions of gas in various
states, from molecular to highly ionized; in starforming
galaxies at $z \approx 2$ the mass in gas is of the same order
of the mass in stars (Cresci et al. 2009; Tacconi et al. 2008,
2010). Such winds would then push out gas from the halo at a
rate
\begin{equation}
\dot {M}_{\rm out}\approx 1000\, \frac{v_{\rm gas}}
{1000\,\mathrm{km~s}^{-1}}\, \left(\frac{R_{\rm
H}}{100\,\mathrm{kpc}}\right)^{-1}\,\frac{M_{\rm
gas}}{10^{11}\, M_{\odot}}~M_{\odot}~\mathrm{yr}^{-1}~,
\end{equation}
where the reference values are for a DM halo with $M_{\rm
H}\approx 2 \times 10^{12}\, M_{\odot}$ formed at $z_{\rm form}
\approx 3$; $v_{\rm gas}$ is the escape velocity from $r\approx
1$ kpc and the gas mass is close to the stellar mass.  As
pointed out by Silk \& Rees (1998) and by Granato et al. (2001,
2004) the energy released by a luminous quasar in its last
$e-$folding time is a factor of $20$ larger than the energy
associated to these winds. On the observational side, hints of
massive outflows from high redshift quasars, consistent with
this scenario, have been reported  (e.g. Simcoe et al. 2006;
Prochaska \& Hennawy 2009). Therefore there are strong reasons
to believe that large gas outflows occurred in high redshift
quasar hosts.

As discussed in \S~\ref{sect:exp}, the removal of a mass in gas
close to the mass in stars destabilizes the mass distribution
in the innermost galaxy regions. In the case of strong quasar
winds the ejection and dynamical timescale are similar
$\tau_{\rm ej}\approx 1-3\, \tau_{\rm dyn}$. Therefore the
effect could be intermediate between those described by
Eq.~(\ref{eq:BS}) and by Eq.~(\ref{eq:adexp}), as found with
numerical simulations by Geyer \& Burkert (2001, cf. their
Fig.~3) and by  Baumgardt \& Kroupa (2007). A basic question is
how long it takes for the stellar structure to readjust to a
new equilibrium. In simulations of star clusters, i.e., without
DM halo, the new equilibrium is reached in $30-50$ initial
crossing times (Geyer \& Burkert 2001; Boily \& Kroupa 2003;
Bastian \& Goodwin 2006). In the hypothesis that the same
number of crossing times are also requested for massive
galaxies, the expected timescale for size evolution would be
$\Delta T_{\rm size}\approx 1.5\times 10^8$ yr.  On the other
hand, specific numerical simulations with high temporal
resolution are needed in order to assess the size evolution
timescale, since the presence of the DM halo, dominating the
potential well at $r\ga R_e$, could slow down the expansion
increasing the time needed to reach the new size and
equilibrium. On the observational side, the duty cycle can be
inferred only by studying the distribution of large samples of
high-$z$ galaxies in the $R_e$ vs. $M_{\star}$ plane.

Since quasar winds follow the time pattern of quasar shining,
the same is expected for the size evolution, except for a delay
by $\Delta T_{\rm size}$. As a consequence, the inverse
hierarchy or downsizing seen in the quasar evolution is
mirrored in the size evolution.

Quasar activity is the main feedback mechanism for more massive
ETGs ($M_{\star}\ga  10^{11}\, M_{\odot}$), while supernova
feedback dominates at $M_{\star}\la 2\times 10^{10}\,
M_{\odot}$ (see Granato et al. 2004; Lapi et al. 2006; Shankar
et al. 2006). Correspondingly, larger size evolution is
expected for larger mass ETGs, while the evolution is
progressively decreasing for lower mass and should be
negligible for $M_{\star}\la 10^{10}\,M_{\odot}$; in addition,
the scatter in size at high $z$ is much wider for more massive
ETGs. Interestingly, Lauer et al. (2007a) and Bernardi et al.
(2007) found that the relationship between effective radius
$R_e$ and luminosity steepens for ETGs brighter than
$M_V\approx -21$, corresponding to a stellar mass
$10^{11}\,M_{\odot}$.

If we assume that the change in size of ETGs is due to minor
dry mergers, we face a couple of problems. The upper limit to
the mass evolution ($M_f/M_i\approx 1.7-2$ since $z\approx
1.5-2$), plus the fact that this happens gradually, implies
that {\it almost all} high redshift massive ETGs must increase
their size at most by a factor $\approx 2.2-3$. While this may
be consistent with the average size evolution, it does not
account for the decreased scatter of the size distribution from
high to low redshifts. In particular, since dry minor mergers
require a long timescale $\approx 10$ Gyr to produce their full
effects, they can not not explain why a significant fraction of
the high-$z$ ETGs are already on the local mass-size
relationship. Moreover, the upper limit on the increase in mass
entails an upper limit of factor of 3 for the size increase
since $z\approx 2$, while at this redshift there are ETGs with
sizes smaller than the local one by factors of $6-10$ .

\subsection{Projected Central Mass evolution}\label{sect:cenmass}

Clearly, the mass within the central regions after the
expansion driven by quasar winds in ETGs, when a new virial
equilibrium is reached {\it with the same mass in stars}, has
to be lower than that of the initial compact structure,
analogously to what happens for stellar clusters (Boily \&
Kroupa 2003; Baumgardt \& Kroupa 2007). To test this
implication against the data we compare the projected mass
within the half-mass radius $R_e(z\ga 1)$ of each passively
evolving galaxies at high redshift with the average projected
mass within the {\it same physical radius} for local ETGs of
the same overall stellar mass.

We stress that this test bypasses the problem of the
reliability for the estimates of the effective radius in high
redshift ETGs, since the mass inside the estimated effective
radius is much less uncertain than the value of the radius
itself. We checked that following the method by Hopkins et al.
(2010), who have illustrated the effect of a limited dynamical
range in surface brightness available for high redshift
galaxies on the estimate of the intrinsic index $n_t$ and of
half-mass radius $R_t$. These authors shifted some of the Virgo
clusters ETGs to $z\approx 2$ and simulated HST observations on
these objects. Specifically, for NGC 4552 shifted to high
redshift and assuming $\Delta\mu\approx 4.5$, Hopkins et al.
(2010) find that the original $n_t=9.22$ is fitted with
$n_f\approx 6$ and $R_f\approx R_t/3$; as a results, the total
luminosity $L_f$ obtained by the fitting procedure is lower
than the `true' one $L_t$ by a factor $1.5$. In the case of NGC
4365 the corresponding luminosity ratio amounts to $1.12$.
However, the luminosity $L_f(R_f)$ inside $R_f$ estimated
through the fit is a good estimate of the true luminosity, with
errors within $20\%$ even in the case of large values of $n_t$.
For instance, in the case of NGC 4552 we get
$L_f(R_f)/L_t(R_f)\approx 1.15$. In conclusion, while the
fitting procedure for high redshift galaxies with large
intrinsic $n$ index underestimates the half-mass radius by a
factor of $3-4$ and the total luminosity by a factor $1.5$, but
yields accurate estimates of the luminosity (and of the mass)
inside the radius $R_f$. By the way, the same holds for the
stellar mass, after proper translation of the luminosity in
mass.

\begin{figure*}
\plotone{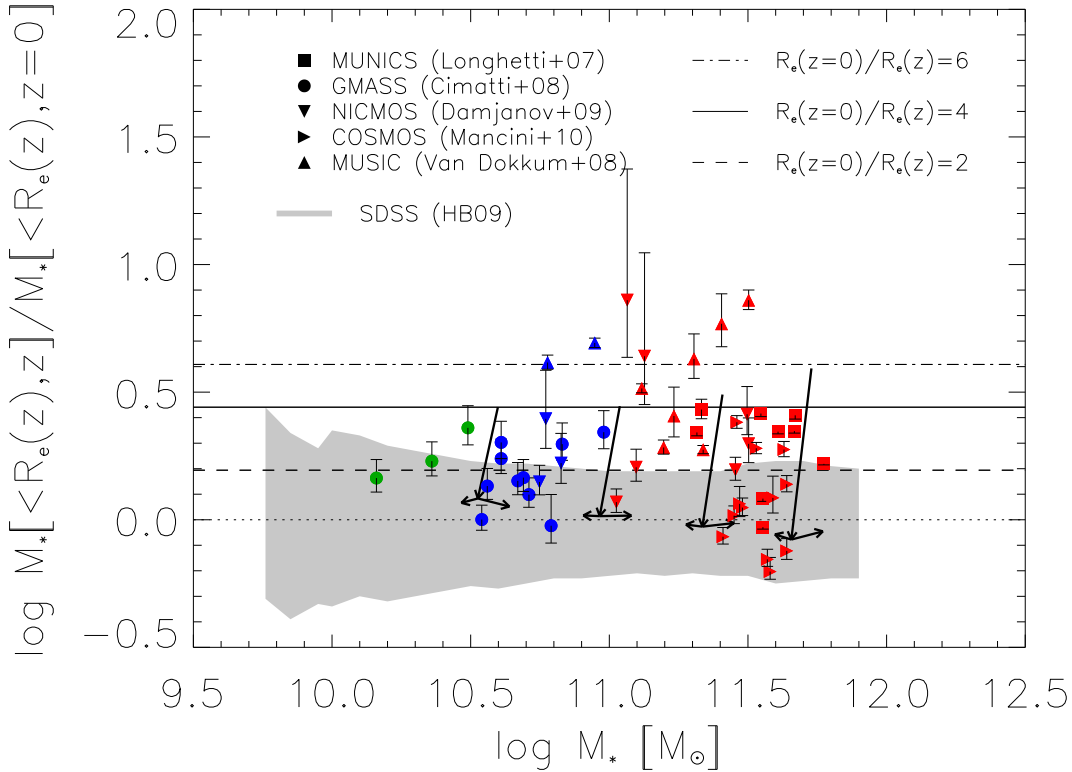} \caption{Ratio between the projected stellar
mass within the estimated half-mass radius $R_e$ for passively
evolving ETG progenitors at $z\ga 1$ and the average local
value within the same physical radius and for the same overall
stellar mass. The data points include observations of
individual passively evolving galaxies with spectroscopic
redshifts $z\ga 1$ by Longhetti et al. (2007), Cimatti et al.
(2008), Damjanov et al. (2009), Mancini et al. (2010), and van
Dokkum et al. (2008). The shaded area shows the distribution of
local SDSS galaxies (Hyde \& Bernardi 2009). The thin lines
illustrate the expected stellar mass ratio for different size
increase from high$-z$ to the present, on adopting a local
S\'ersic index $n=4$ (see \S~\ref{sect:cenmass} for more
details). As in Fig.~2, thick lines with
arrows illustrate typical evolutionary tracks of massive
galaxies according to our reference model (with
$f_{\sigma}=1.5$ and $z_{\rm form}=3$).}\label{fig:iratio}
\end{figure*}

In Fig.~4 we plot the ratio between the projected mass of high
redshift and local ETGs \emph{within the same physical radius},
namely, the half-luminosity radius of high-redshift ETGs. In
detail, since for high redshift ETGs
$M_{\star}(<R_e(z),z)=M_{\star, \rm tot}/2$, the ratio comes to
\begin{equation}
\frac {M_{\star}(<R_e(z),z=0)}{M_{\star}(<R_e(z),z)}=2\,{\Gamma\left[2\,n,
b_n\,\left({R_e(z)\over \langle R_e(z=0)\rangle
}\right)^{1/n}\right]}~,
\end{equation}
where $\Gamma$ is the (normalized) incomplete Gamma function.
Thus if the total mass does not change significantly, the
quantity $M_{\star}(<R_e(z),z)/M_{\star}(<R_e(z),z=0)$ depends
only on the ratio $r=R_e(z)/\langle R_e(z=0)\rangle$ and on the
final S\'{e}rsic index $n$.

To plot the data points under the hypothesis of no mass
evolution, we compute the ratio $r$ using the observed
$R_e(z)$, and exploiting the observed stellar mass $M_{\star}$
to derive $\langle R_e(z=0)\rangle $ from the local
$R_e-M_{\star}$ relation presented in Fig.~2. The horizontal
lines has been computed for three values of $r$; for the sake
of definiteness we adopt $n=4$. The thick lines with arrows
illustrate the evolutionary track of massive galaxies according
to our reference model with $f_{\sigma}=1.5$ and $z_{\rm
form}=3$; these are found to be in encouraging agreement with
the distribution of data points.

In the case of local ETGs we can define a ratio $r'\equiv
R_e(z=0)/ \langle R_e(z=0) \rangle $ and compute the analogous
of the mass ratio defined by Eq.~(14). The shaded area,
containing $65\%$ of local ETGs, illustrates the uncertainty of
data points and of the horizontal lines associated to the
assumption of an average half radius $\langle R_e(z=0)\rangle$.

Data points in Fig.~4 show that a significant fraction of high
redshift passively evolving galaxies exhibit stellar mass
inside their inferred half-mass radius larger by a factor $2-6$
than the mass of their local counterparts within the same
physical radius.

Keeping mass and structural index $n=4$ constant, larger mass
ratios can be obtained increasing the half luminosity radius by
a factor from $2$ to $6$ (cf. horizontal lines). If we allow
the structural index $n$ to vary by one, as suggested by the
simulations of Naab et al. (2009), the change is minimal; even
structural changes from $n=4$ to $n=8$ still would require a
size increase by a factor of $\approx 6$ in order to explain
galaxies with the largest mass ratio.

This result suggests that the mass inside this physical radius
has on the average decreased and disfavors mechanisms that
increase the size by only adding stars in the outer regions of
the ETGs. We notice that our argument involves a significant
fraction of the total galaxy mass. Contrariwise, the comparison
of stellar surface density profiles within $R_e/50$ as
performed by Hopkins et al. (2009a) refers only to a tiny
fraction of the mass.  It is interesting to note that Naab et
al. (2009) find in their simulations that the dynamical
friction is able to decrease the total mass inside $\la 1$ kpc.

On the same line, massive ETGs with their large sizes, steeper
correlation between effective radii and mass and large
S\'{e}rsic index (Lauer et al. 2007b; Kormendy et al. 2009)
clearly stand as representative cases of galaxies which
experienced robust puffing up by quasar feedback.  Moreover,
the correlation of the central BH mass with S\'{e}rsic index
$n$ (Graham et al. 2003; Graham \& Driver 2007) for massive
ETGs is consistent with the hypothesis that the strong feedback
from the most massive BHs has led to a substantial increase of
$n$.

\begin{figure*}
\plotone{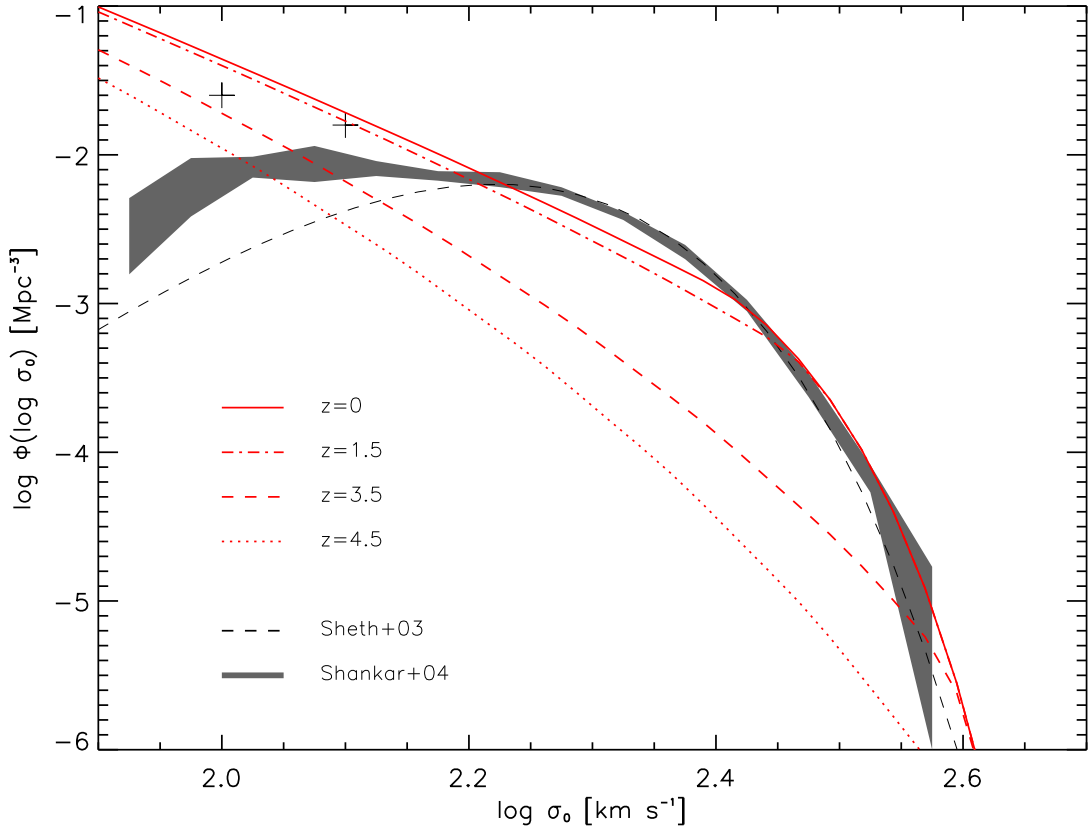}
\caption{Velocity distribution function. Our results at
different redshifts are
compared with the observational estimates of the local velocity
distribution function by Sheth et al. (2003) and Shankar et al.
(2004).}\label{fig:vdf}
\end{figure*}

\subsection{Velocity dispersion evolution}\label{sect:VDF}

As mentioned in \S\,\ref{sect:vdisp}, velocity dispersion has
so far been determined only for a few individual high-$z$
spheroidal galaxies. The galaxy GMASS 2470 at $z\approx 1.4$
(Cappellari et al. 2009) and galaxy $\#$250425 at $z=1.82$
(Onodera et al. 2010) are already close to the local value for
their mass, while $1255-0$ at $z\approx 2.2$ has a best fit
velocity dispersion significantly higher than the most massive
local galaxies. Thus in the two first cases evolution should
have occurred before the cosmic time corresponding to the
observed redshift, whereas significant evolution in size and
velocity dispersion has to occur for the higher redshift
galaxy. The studies of velocity dispersions by Cenarro \&
Trujillo (2009) and Cappellari et al. (2009), based on stacked
spectra, suggest that velocity dispersion evolution is on the
average needed.

An interesting hint on size and velocity dispersion average
evolution can be derived by studying the velocity dispersion
distribution (VDF) of local ETGs (Sheth et al. 2003), following
the approach by Cirasuolo et al. (2005). From
Eqs.~(\ref{eq:sigmaVH}) and (\ref{eq:BS}), taking $n_i=3$ and
$n_f=5$, we find, for the typical values of the parameters
discusses above ($f_\sigma=1.5$, $r=4$), the relation
$\sigma_{0,f}= 0.6\, V_{\rm H}$; this associates to each
forming halo the final velocity dispersion of the most massive
hosted galaxy. By combining it with the formation rate of halos
(see Appendix A) for redshifts $z\ga 1$ we get the distribution
function of the stellar velocity dispersions (VDF). From
Fig.~\ref{fig:vdf} it is apparent that the predicted VDF is in
good agreement with the observational estimate by Sheth et al.
(2003). We stress that the observed VDF highly constrains the
global history of DM galaxy halos and of their stellar content,
and therefore is an important benchmark for models of ETG
formation (see also Loeb \& Peebles 2003).

A change of the slope of the relationship between luminosity
and velocity dispersion at low luminosity has been claimed by
several authors (Shankar et al. 2006; Lauer et al. 2007a). In
particular, Lauer et al. (2007a) show that the change in slope
occurs at about the same luminosity $M_V\approx -21$, where the
slope of the size vs. luminosity relation also changes (see
\S\,6.2).  Once more this feature occurs at the transition from
supernova-dominated to quasar-dominated feedback regime.

An additional relevant aspect is related to the $M_{\bullet}$
vs. $\sigma_{0}$ and $M_{\bullet}$ vs. $M_{\star}$
correlations. The Granato et al. (2004) model, which we take as
a reference, predicts that the mass in stars formed before the
quasar shining is strictly related to the mass of the central
BH, since the growth of the reservoir, which eventually
furnishes the mass to the BH, is strictly proportional to the
starforming activity  (cf. Eq.~[A20]). Marconi \& Hunt (2003)
and H\"{a}ring \& Rix (2004) pointed out that the $M_{\bullet}$
vs. $M_{\star}$ relation for ETGs and bulges exhibits a scatter
comparable with that in the $M_{\bullet}$ vs. $\sigma_{0}$
relation.

In this context it is interesting to mention that, according to
Lauer et al. (2007a), the extrapolation of the $\sigma_0$ vs.
$M_{\bullet}$ relationship holding at low mass to higher mass
galaxies would predict BH masses smaller than those inferred
from the stellar mass. This is expected if the velocity
dispersion decreased more significantly for higher mass
galaxies, hosting more massive BHs and subject to stronger
quasar feedback.

\section{Summary and conclusions}

The half-luminosity radius of high redshift passively evolving
massive galaxies is observed to be on the average significantly
smaller than that of their local counterparts with the same
stellar mass, but in agreement with theoretical predictions
based on the largely accepted assumption that most of the stars
have been formed during dissipative collapse of cold gas.
However, observations also show that the size distribution of
high redshift ETG progenitors is broader than the corresponding
distribution for local ETGs. While a significant fraction of
massive high redshift ETGs already exhibit sizes as large as
those of their local counterparts with the same mass, for a
bunch of ETGs the size has to increase by a factor of $5-10$ to
match the local half mass radius. Though still scanty, the
available data on velocity dispersions are suggestive of a
correspondingly large scatter of the ratios between high-$z$
and local values at fixed stellar mass.

The analysis of several data sets, discussed in
\S\,\ref{sect:sizevol}, and notably of the large sample by
Maier et al. (2009) with spectroscopic redshifts, strongly
suggests that most of the size evolution occurs at $z \ga 1$,
while at $z \la 1$ sizes increase by no more than $40\%$.
Moreover, a large fraction of high-$z$ passively evolving
galaxies have projected stellar mass within their effective
radii a factor of $2$ larger than those of local ETGs with the
same stellar mass, within the same physical radius (see
\S~\ref{sect:sizev}).

All the above results are easily accounted for if most of the
size evolution is due to a puffing up driven by the rapid
expulsion of large amounts of mass, as proposed by Fan et al.
(2008). That most of the baryons initially associated to the DM
halo have to be expelled is strongly indicated by the fact that
the baryon to DM mass ratio in galaxies is much smaller than
the cosmic value. The quasar-driven winds advocated by Fan et
al. (2008) occur in the most massive galaxies, while below
$M_\star \approx 2\times 10^{10}\,M_\odot$ the dominant energy
input into the interstellar medium comes from supernova
explosions which induce a slower mass loss. The Fan et al.
(2008) model therefore predicts a milder size evolution for the
less massive spheroidal galaxies, while the size evolution of
the more massive galaxies should parallel the quasar evolution,
with a delay of about $0.5-1$ Gyr. The dichotomy between low-
and high-mass galaxies, i.e., between supernova and quasar
driven feedback, is mirrored in the increase of S\'{e}rsic
index with stellar mass, in the flattening of the total mass
vs. velocity dispersion relation toward massive galaxies, and
in corresponding steepening of the correlation between
effective radii and stellar mass.

The alternative explanation invoking minor mergers faces a
couple of difficulties (see also Nipoti et al. 2009a,b). The
analysis of the mass function evolution shows that, under the
hypothesis of pure mass evolution, the upper limits to the mass
increase are a factor $\approx 2$ and a factor $\approx 1.7$
since $z\approx 2$ and $z\approx 1.5$, respectively. Also the
increase is expected to be gradual and rather uniform, so that
practically all galaxies undergo the same mass increase. As a
consequence {\it almost all} high redshift massive ETGs must
evolve by  a factor $\la 2.2-3$. While this upper limit may be
consistent with the average evolution of the size, the model
does not account for the substantially broader size
distribution, for given stellar mass, at high, compared to low,
redshifts. In particular, since dry minor mergers require a
long timescale $\approx 10$ Gyr to produce their full effects,
they can not not explain why a significant fraction of the
high-$z$ ETGs are already on the local mass-size relationship.
Moreover, the upper limit in mass entails a factor of $\la 3$
in size evolution since $z\approx 2$, while at the same
redshift there are ETGs with half mass undersized by a factor
$6-10$. On the positive side, for the minor merger scenario,
the simulations by Naab et al. (2009) show that dynamical
friction is able to remove part of the mass from the central
regions in line with what suggested by observations (see
Fig.~4).

The virial theorem tells us that the velocity dispersion scales
as $\sigma^2 \propto S_D\, M/r$, where $S_D$ is a structure
factor defined by Prugniel \& Simien (1997) in the case of a
S\'{e}rsic profile. Since the rapid loss of a large mass
fraction destabilizes the mass distribution, it may be expected
that the final equilibrium configuration differs from the
initial one. In fact, the data by van Dokkum et al. (2010) and
simulations indicate that the S\'{e}rsic index of local
galaxies is, on average, higher than for high-$z$ galaxies. If
so, the variation of $S_D$ partially compensates the effect on
$\sigma$ of the size increase. Although measurements of the
kinematic properties of high-$z$ galaxies are scarce, a
velocity dispersion evolution compatible with the expansion
scenario is indicated (see \S~\ref{sect:VDF}).

\begin{acknowledgements}
This research has been partially supported by ASI under
contract I/016/07/0 `COFIS'. A.B. is grateful for support
provided by NASA grants LTSA-NNG06GC19G and ADP/ NNX09AD02G. We
thank the referee for comments and suggestions helpful toward
improving our presentation. We acknowledge useful discussions
with F. Shankar. A.L. thanks SISSA and INAF-OATS for warm
hospitality.
\end{acknowledgements}

\appendix

\section{Overview of our reference model}

In recent years we developed a model of galaxy formation with
focus on the evolution of baryons within protogalactic
spheroids. Baryons have been followed through simple physical
recipes emphasizing the effects of the collapse and cooling and
of the energy fed back to the intragalactic gas by supernova
(SN) explosions and by accretion onto the nuclear supermassive
black holes (BHs; see Granato et al. 2001, 2004; Lapi et al.
2006, 2008; Mao et al. 2007; Fan et al. 2008). The main
motivation was to enlighten the relevant physical processes
shaping galaxy formation, to keep calculations easily
reproducible and to suggest which processes should be
implemented in the much more complex and much less reproducible
numerical simulations.

The model transparently shows how physical processes acting on
the baryons speeds up the formation of more massive galaxies.
As a result, although the DM assembly follows a bottom-up
hierarchy, galaxies and their active nuclei evolve in a way
that appears opposite to the hierarchy in DM, following a
pattern that we named Antihierarchical Baryon Collapse (ABC).
We notice that it fully corresponds from the observational
point of view to the so called downsizing.

We defer the interested reader to the above papers for a full
account of the physical justification and a detailed
description of the model, with appropriate acknowledgment of
previous work. Here we present a short summary of its main
features, and provide useful analytic approximations for
quantities of relevance in this context.

\subsection{DM sector}

As for the treatment of the DM in galaxies, the model follows
the standard hierarchical clustering framework, and takes into
account the outcomes of recent intensive high-resolution
$N-$body simulations of halo formation in a cosmological
context (see Zhao et al. 2003; Diemand et al. 2007; Hoffmann et
al. 2007; Ascasibar \& Gottloeber 2008). In these studies, two
distinct phases in the growth of DM halos have been recognized:
an early fast collapse, and a later slow accretion phase.
During the early collapse, a substantial mass is gathered
through major mergers, which effectively reconfigure the
gravitational potential wells and cause the collisionless DM
particles to undergo dynamical relaxation and isotropization
(Lapi \& Cavaliere 2009). During the later phase, moderate
amounts of mass, around $20-50\%$, are slowly accreted mainly
onto the halo outskirts, little affecting the inner structure
and potential, but quiescently rescaling upward the overall
halo size. From the baryon point of view, the early phase ---
our main interest here --- supports the dissipationless
collapse of baryons to originate a spheroidal structure
dominated by random motions (see also Cook et al. 2009).

Halos harboring a massive elliptical galaxy once created, even
at high redshift, are rarely destroyed, while at low redshifts
they are possibly incorporated within galaxy groups and
clusters. Thus at $z\ga 1$, the halo formation rate can be
reasonably well approximated by the positive term in the
derivative of the halo mass function with respect to cosmic
time (e.g., Haehnelt \& Rees 1993; Sasaki 1994). The halo mass
function derived from numerical simulations (e.g., Jenkins et
al. 2001) is well fit by the Sheth \& Tormen (1999, 2002)
formula, that improves over the original Press \& Schechter
(1974) expression (well known to under-predict by a large
factor the massive halo abundance at high redshift). Adopting
the Sheth \& Tormen (1999) mass function
$N_{\mathrm{ST}}(M_{\rm H}, t)$, the formation rate of DM halos
is given by
\begin{equation}
{\mathrm{d}^2\, N_{\mathrm{ST}}\over \mathrm{d} M_{\rm H}\, \mathrm{d}
t}=\left[{a\, \delta_c(t)\over \sigma^2(M_{\rm H})}+{2\, p\over
\delta_c(t)}\, {\sigma^{2\,p}(M_{\rm H})\over \sigma^{2\,p}(M_{\rm H})+ a^p\,
\delta_c^{2\,p}(t)}\right]\, \left|{\mathrm{d}\delta_c\over
\mathrm{d} t}\right|\, N_{\mathrm{ST}}(M_{\rm H}, t)~;
\end{equation}
here $a=0.707$ and $p=0.3$ are constants obtained by comparison
with $N$-body simulations, $\sigma(M_{\rm H})$ is the mass
variance of the primordial perturbation field computed from the
Bardeen et al. (1986) power spectrum with correction for
baryons (Sugiyama 1995) and normalized to $\sigma_8\approx 0.8$
on a scale of $8\,h^{-1}$ Mpc, and $\delta_c(t)$ is the
critical threshold for collapse extrapolated from the linear
perturbation theory.

As for the density distribution of DM halos we adopt as a
reference the profile proposed by Navarro et al. (1997) and
characterized by a scale radius $r_s$ and by the ratio of the
virial to the scale radius $c=R_{\rm H}/r_{s}$, the
concentration parameter, with typical values around $4$ at halo
formation (e.g., Zhao et al. 2003). The halo circular velocity
$V_{\rm H}=(G\, M_{\rm H}/R_{\rm H})^{1/2}$ characterizes the
DM potential well; the associated velocity dispersion is
$\sigma_{\rm DM}=f(c)^{1/2}\, V_{\rm H}$, where $f(c)\approx
2/3 +(c/21.5)^{0.7}$ is a weak function of the concentration
parameter of order $1$ (see Mo et al. 1998).

\subsection{Baryonic sector}

During the fast collapse phase, a rapid sequence of major
mergers build up a DM halo of mass $M_{\rm H}$. At that time a
mass $M_{\rm inf}=f_b\,M_{\rm H}$ of baryonic matter, in cosmic
proportion $f_b\approx 0.2$ with the DM, is shock heated to the
virial temperature by falling into the forming DM gravitational
potential well. This hot gas cools and flows toward the central
region at a rate
\begin{equation}
\dot{M}_{\mathrm{cond}}={M_{\mathrm{inf}}\over t_{\mathrm{cond}}}
\end{equation}
over the {\it condensation} timescale
$t_{\mathrm{cond}}=\max[t_{\mathrm{cool}}(R_{\rm
H}),t_{\mathrm{dyn}}(R_{\rm H})]$, namely, the maximum between
the dynamical and the cooling time at the halo virial radius
$R_{\rm H}$. When computing the cooling time, a clumping factor
in the gas $\mathcal{C}\ga$ a few, as suggested by numerical
simulations (e.g., Gnedin \& Ostriker 1997; Iliev et al. 2006),
implies $t_{\mathrm{cool}}(R_{\rm H})\la
t_{\mathrm{dyn}}(R_{\rm H})$ on relevant galaxy scales at $z\ga
1$.

We recall that the star formation in galaxy halos is a quite
inefficient process (see Fig.~A1), since only a minor fraction
$10-20\%$ of the available baryons are cycled through stars in
more massive halos and the fraction is rapidly decreasing with
decreasing halo mass. As a result, the present-day cosmic mass
density in stars is only a few percent of the mass density in
baryons (e.g., Shankar et al. 2006). Thus the formation of the
most massive galaxy in a halo is a process that involves only a
fraction smaller than $20-30\%$ of its original baryons and DM.
It is natural to assume that the material is rapidly put
together by a few major mergers in the central regions of the
halo. In these mergers the angular momentum decays on a
dynamical friction timescale $t_{\mathrm{DF}}\approx 0.2\,
(\xi/\ln{\xi})\, t_{\mathrm{dyn}}$, where $\xi\equiv M_{\rm
H}/M_{\rm c}$ holds in terms of typical cloud mass $M_{\rm c}$
involved in major mergers (e.g. Mo \& Mao 2004); these are very
frequent at high redshift and in the central regions of halos
during the fast collapse phase of DM evolution, implying $\xi
\sim$ a few and hence a short $t_{\mathrm{DF}}$. Thus the
effects of angular momentum can be neglected.

The model also assumes that quasar (QSO) activity removes not
only cold gas from the galaxy, but also hot gas from the halo
through winds at a rate $\dot{M}_{\mathrm{inf}}^{QSO}$, to be
quantitatively discussed next; the equation for the diffuse hot
gas is then
\begin{equation}
\dot{M}_{\mathrm{inf}}=
-\dot{M}_{\mathrm{cond}}-\dot{M}_{\mathrm{inf}}^{QSO}~.
\end{equation}

The cold gas piled up by the cooling of hot gas, is partially
consumed by star formation ($\dot{M}_{\star}$), and partially
removed to a warm/hot phase endowed with long cooling time by
the energy feedback from SNae ($\dot{M}_{\mathrm{cold}}^{SN}$)
and QSO activity ($\dot{M}_{\mathrm{cold}}^{QSO}$):
\begin{equation}\label{eq:Mcold}
\dot{M}_{\mathrm{cold}} = \dot{M}_{\mathrm{cond}}-
[1-\mathcal{R}(t)]\dot{M}_{\star} -
\dot{M}_{\mathrm{cold}}^{SN}-\dot{M}_{\mathrm{cold}}^{QSO}~,\\
\end{equation}
where $\mathcal{R}(t)$ is the fraction of gas restituted to the
cold component by the evolved stars. It depends on time
(particularly soon after the onset of star formation) and on
the assumed initial mass function (IMF). We adopt for reference
a pseudo-Chabrier IMF of shape $\phi(m_\star) = m_\star^{-x}$
with $x=1.4$ for $0.1\leqslant m_{\star}\leqslant 1\,M_{\odot}$
and $x=2.35$ for $m_{\star}>1\,M_{\odot}$. The  often used
approximation of instantaneous recycling implies
$\mathcal{R}\approx 0.54$  (for a Salpeter IMF one has
$\mathcal{R}\approx 0.3$). The mass of cold baryons that is
going to be accreted onto the central supermassive BH is small
enough to be neglected in the above equation.

Stars are formed at a rate
\begin{equation}
\dot{M_{\star}}=\int
\frac{\mathrm{d}M_{\mathrm{cold}}}{\max[t_{\mathrm{cool}},t_{\mathrm{dyn}}]}
\approx \frac {M_{\mathrm{cold}}}{t_{\star}}~,
\end{equation}
where now $t_{\mathrm{cool}}$ and $t_{\mathrm{dyn}}$ refer to a
mass shell $\mathrm{d}M_{\mathrm{cold}}$, and $t_{\star}$ is
the star formation timescale averaged over the mass
distribution. Star formation promotes the gathering of some
cool gas into a low-angular-momentum reservoir around the
central supermassive BH. A viable mechanism for this process is
the radiation drag (see discussion by Umemura 2001; Kawakatu \&
Umemura 2002; Kawakatu et al. 2003), which has the nice feature
of predicting a mass transfer rate to the reservoir
proportional to the SFR:
\begin{equation}
\dot{M}_{\rm inflow} \approx \frac{L_{\star}}{c^2}\,\left
(1-e^{-\tau_{\rm RD}}\right)\approx \alpha_{\rm RD}\times
10^{-3}\, \dot{M}_{\star}\, \left(1-e^{-\tau_{\rm RD}}\right)~;
\end{equation}
the constant of proportionality $\alpha_{\rm RD}\sim 1-3$ can
be fixed to produce a good match to the correlation between the
spheroid and the supermassive BH masses observed in the local
Universe, while the quantity
\begin{equation}
\tau_{\rm RD}(t) = \tau_{\rm RD}^0\,{Z(t)\over
Z_{\odot}}\, {M_{\rm cold}(t)\over 10^{12}\,
M_{\odot}}\, \left({M_{\rm H}\over 10^{13}\,
M_{\odot}}\right)^{-2/3}~
\end{equation}
represents the effective optical depth of the gas clouds in
terms of the normalization parameter $\tau_{\rm RD}^0\sim 1-2$
(for more details, see the discussion around Eqs.~[14] to [17]
in Granato et al. 2004).

Eventually, the gas stored in the reservoir accretes on to the
BH powering the nuclear activity; usually, plenty of material
is supplied to the BH, so that the latter can accrete close to
the Eddington limit
\begin{equation}
\dot{M}_{\bullet}=\lambda_{\rm Edd}\,{1-\epsilon\over
\epsilon}\,{M_{\bullet}\over t_{\rm Edd}}
\end{equation}
and grows almost exponentially from a seed of
$10^2\,M_{\odot}$; the $e-$folding time involves the Eddington
timescale $t_{\rm Edd}\approx 4\times 10^8$ yr, the radiative
efficiency $\epsilon\approx 0.1$, and the actual Eddington
ratio $\lambda_{\rm Edd}\sim 0.3-3$. The reservoir mass
variation is ruled by the balance between the inflow due to
radiation drag and the accretion onto the BH
\begin{equation}
\dot{M}_{\rm res} = \dot{M}_{\rm inflow} - \dot{M}_{\bullet}~,
\end{equation}
between the inflow due to radiation drag and the accretion onto
the BH.

The energy fed back to the gas by SN explosions and BH activity
regulates the ongoing star formation and BH growth. The two
feedback processes have very different dependencies on halo
mass and on galaxy age. The feedback due to SN explosions
removes the starforming gas at a rate
\begin{equation}\label{eq|mdotsn}
\dot{M}_{\mathrm{cold}}^{SN} = \beta_{SN}\, \dot{M}_{\star}~,
\end{equation}
where the efficiency of gas removal
\begin{equation}\label{eq|betasn}
\beta_{SN}=\frac {N_{SN}\,
\epsilon_{SN}\,E_{SN}}{E_{\mathrm{bind}}}\approx 0.6
\left(\frac{N_{SN}}{8\times 10^{-3}/ M_{\odot}}\right)
\left(\frac{\epsilon_{SN}}{0.05}\right)
\left(\frac{E_{SN}}{10^{51}\mathrm{erg}}\right)
\left(\frac{M_{\rm H}}{10^{12} M_{\odot}}\right)^{-2/3} \left(
\frac{1+z}{4}\right)^{-1}
\end{equation}
depends on the number of SNae per unit solar mass of condensed
stars $N_{SN}$ (of order $1.4 \times 10^{-2}$ for the Chabrier
IMF), on the energy per SN available to remove the cold gas
$\epsilon_{SN}\,E_{SN}$, and on the specific binding energy of
the gas within the DM halo, $E_{\mathrm{bind}}$. Following Zhao
et al. (2003) and Mo \& Mao (2004), the latter quantity has
been estimated for $z\ga 1$ as
$E_{\mathrm{bind}}=V_{\mathrm{H}}^{2}\,
f(c)\,(1+f_{\mathrm{b,i}})/2\approx 3.2\times 10^{14}\, (M_{\rm
H}/10^{12}\, M_{\odot})^{2/3}\, [(1+z)/4]\,
{\mathrm{cm}^2~\mathrm{s}^{-2}}$

In Granato et al. (2004) it is assumed that the QSO feedback
acts both on the cold as well as on the infalling gas,
unbinding them from the DM halo potential well at a rate
$\dot{M}_{\rm inf, cold}^{\rm QSO}=\dot{M}_{\rm wind}\times
M_{\rm inf, cold}/(M_{\rm inf}+M_{\rm cold})$ proportional to
the corresponding mass fraction and to the wind mass outflow
rate
\begin{equation}
\dot{M}_{\rm wind}= \frac{L_{QSO}}{E_{\rm bind}}~,
\end{equation}
with
\begin{equation}
L_{QSO}\approx 2\times 10^{44}\, \epsilon _{QSO} \left(\frac
{M_{\bullet}}{10^8\ M_{\odot}}\right)^{3/2}~\mathrm{erg~s}^{-1}~.
\end{equation}
Here $\epsilon_{QSO}=(f_h/0.5)\, (f_c/0.1)\, N_{23}$ is the
strength of QSO feedback, expected to be close to unity; $f_h$
parameterizes the efficiency of energy transfer from winds
generated close to the accretion disc to the general
interstellar medium, $f_c$ is the covering factor of such winds
and $N_{23}$ is the hydrogen column density toward the BH in
units of $10^{23}$ cm$^{-2}$ (cf. Eqs.~[29], [30] and [31] in
Granato et al. 2004). With $\epsilon_{\rm QSO}\approx 1.3$ the
bright end of the galaxy luminosity function is reproduced
(Lapi et al. 2006).

As a consequence, the QSO feedback grows exponentially during
the early phases of galaxy evolution, following the exponential
growth of the supermassive BH mass. It is is negligible in the
first $0.5$ Gyr in all halos, but abruptly becomes dominant in
DM halos more massive than $10^{12}\, M_{\odot}$. Eventually,
in these systems most of the gas becomes unbound from the
potential well of the galaxy halo, so that star formation and
BH activity itself come to an end on a timescale which is
shorter for more massive galaxies.

Indeed, the positive feedback on BH growth caused by star
formation, in cooperation with the immediate and negative
feedback of SN, and the abrupt and dramatic effect of QSO
feedback, are able to reverse the formation sequence of the
baryonic component of galaxies compared to that of DM halos:
the star formation and the buildup of central BHs are completed
more rapidly in the more massive halos, thus accounting for the
phenomenon now commonly referred to as \emph{downsizing}.

The model yields as outputs the time evolutions of stars and
gas (including metals) components of the galaxies and of the
associated active galactic nuclei. When the star formation, gas
abundance and the chemical evolution history of a galaxy within
a DM halo of given mass have been computed, the dust abundance
can be inferred (e.g., Mao et al. 2007). Then the spectral
energy distribution as a function of time from extreme-UV to
radio frequencies can be obtained through spectrophotometric
codes, such as GRASIL, which includes a tretment of dust
reprocessing (Silva et al. 1998; Schurer et al. 2009). Coupling
these results with the DM halo formation rate, we can obtain
the statistics of galaxies and supermassive BHs/QSOs as a
function of cosmic time, in different observational bands.

\subsection{Analytic approximations}

By analyzing the results of the numerical solution for the full
set of equations given above, it is apparent that in massive
galaxies the term of QSO feedback is important only during the
final stage of BH growth, around $2-3$ $e-$folding times
(approximately $10^8$ yr) before the peak of QSO luminosity,
when the energy discharged by the QSO is so powerful to unbind
most of the residual gas, quenching both star formation and
further accretion onto the supermassive BH. On the other hand,
in less massive galaxies the central BH mass and the associated
accretion is not able to stop star formation, which lasts for
several Gyrs. The duration of the star formation $\Delta
t_{\mathrm{burst}}$ can be approximated by a simple analytical
form
\begin{equation}
\Delta t_{\mathrm{burst}}\approx 6\times 10^8 \,
\left(\frac{1+z}{4}\right)^{-1.5}\, \mathcal{F}\left(\frac
{M_{\rm H}}{10^{12}\, M_{\odot}}\right)~~\mathrm{yr}~,
\end{equation}
where $\mathcal{F}(x)=1$ for $x\ga 1$ and
$\mathcal{F}(x)=x^{-1}$ for $x\la 1$. A good approximation for
the star formation history in massive galaxies is obtained by
neglecting the QSO feedback effect in Eqs.~(A3) and (A4) and by
abruptly stopping star formation and accretion onto the central
BH after $\Delta t_{\mathrm{burst}}$ since halo formation.

Then Eqs.~(A3) and (A4) can be easily solved, with the outcome
that the infalling mass declines exponentially as
\begin{equation}
M_{\mathrm{inf}}(t) = M_{\rm
b,i}\,\mathrm{e}^{-t/t_{\mathrm{cond}}}~,
\end{equation}
where we assume $M_{\rm b,i}=f_{b}\,M_{\rm H}$, with
$f_{b}=\Omega_{b}/\Omega_{\rm DM}\approx 0.2$. The cold gas
mass evolves according to
\begin{equation}
M_{\mathrm{cold}}(t)=\frac
{f_{b}M_{\rm H}}{s\gamma-1}\, \left[
\mathrm{e}^{-t/t_{\mathrm{cond}}}- \mathrm{e}^{-s\,\gamma\,
t/t_{\mathrm{cond}}}\right]~;
\end{equation}
here we have introduces the shorthand $\gamma\equiv
1-\mathcal{R}+\beta_{\rm SN}$. The quantity $s\equiv
t_{\mathrm{cond}}/t_{\mathrm{\star}}$ is the ratio between the
timescale for the large-scale infall estimated at the virial
radius and the star formation timescale in the central region;
it corresponds to $s\approx 5$, both for an isothermal or NFW
density profile with standard value of the concentration
parameter $c\approx 4$ at halo formation. We notice that the
dependence on the fraction of gas restituted by the stars is
quite weak and that the value obtained by the hypothesis of
instantaneous recycling can be used.

\begin{figure*}
\plotone{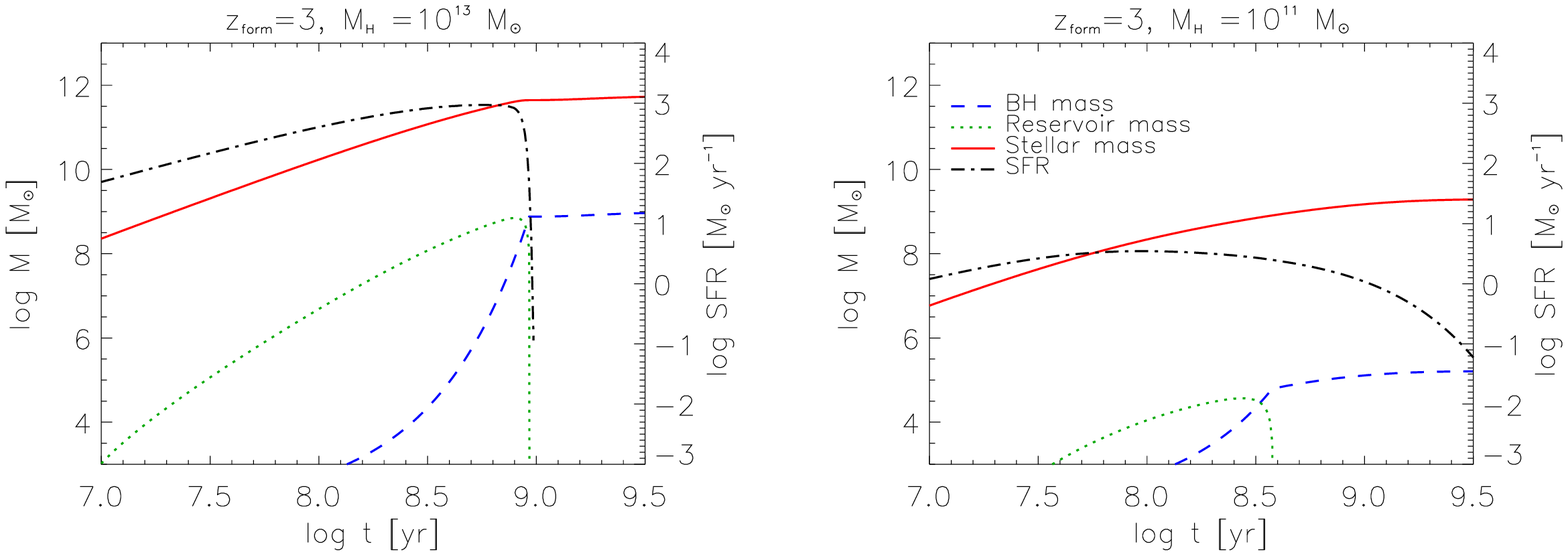} \caption{Evolution with galactic age of the
stellar, reservoir, BH masses (left axis scale) and of the SFR
(right axis scale) in our reference model (Granato et al. 2004)
for galaxy halos with mass $10^{13}\, M_{\odot}$ and $10^{11}\,
M_{\odot}$  formed at $z_{\rm form}=3$.}
\end{figure*}

The corresponding stellar mass reads:
\begin{equation}
M_{\star}(t)=\frac {s\,f_{b}M_{\rm H}}{s\gamma-1}\, \left[1-
\mathrm{e}^{-t/t_{\mathrm{cond}}}-{1\over
s\gamma}\,\left(1-\mathrm{e}^{-s\,\gamma\,
t/t_{\mathrm{cond}}}\right)\right]~;
\end{equation}
the mass of all stars formed during the main episode of star
formation is then $M_{\star}^{\rm f}\approx M_{\star}(\Delta
t_{\mathrm{burst}})$.

In Eqs.~(A13), (A14), and (A16) the condensation timescale is
well approximated by
\begin{equation}
t_{\mathrm{cond}}\approx 9\times 10^8\,
\left(\frac{1+z}{4}\right)^{-1.5}\, \left(\frac {M_{\rm H}}{10^{12}\,
M_{\odot}}\right)^{0.2}~~\mathrm{yr}~.
\end{equation}
The scaling with redshift and mass mainly reflects the behavior
of the dynamical/cooling time, while a mild dependence on
$M_{\rm H}$ is also related to the impact of the energy
feedback from the QSO on the infalling gas, which is stronger
for more massive halos hosting more massive BH.

A good approximation for $f_{\rm gas}$, i.e. the ratio between
the stellar and gaseous mass immediately before the gas is
swept away by the QSO feedback, can be  obtained using
Eqs.~(A16) and (A17). The gas mass includes both the cold gas
and the gas restituted by the stars. The latter is estimated as
$\mathcal{R}\,M_{\star}$ with $\mathcal{R}=0.3$, which
corresponds to the gas returned by the stars about $0.1$ Gyr
after a burst of star formation with a Chabrier's IMF. We find
that the results are well approximated by $f_{\rm gas}\approx
(M_{\star}/10^{11}\, M_{\odot})^{0.2}$.

For $M_{\rm H}\ga 3\times 10^{11}\,M_{\odot}$, corresponding to
$M_{\star}\ga 10^{10}\,M_{\odot}$, the dependence on halo mass
and formation redshift of ratio $m$ between the halo mass and
the surviving stellar mass (i.e. the present day stellar mass
fraction) can be approximated as  $m\approx 25\, (M_{\rm
H}/10^{12}\, M_{\odot})^{0.1}\, [(1+z_{\rm form})/4]^{-0.25}$.
At lower masses $m$ increases rapidly with decreasing
$M_{\star}$ (see Shankar et al. 2006).

Finally, considering that all the mass flowed into the
reservoir is eventually accreted by the BH and neglecting the
mass of the seed BH ($M_{\bullet}^0 \approx 10^2\, M_{\odot}$),
one can write the relic BH mass as function of the overall mass
in stars formed during the star forming phase $\Delta t_{\rm
burst}$ as
\begin{equation}
M_{\bullet}^{\rm f}\approx  \alpha_{\rm RD}\times 10^{-3}\,
M_{\star}^{\rm f}\, (1-{\rm e}^{-\langle\tau_{\rm RD}\rangle})~.
\end{equation}
The time average of the optical depth $\langle\tau_{\rm
RD}\rangle$ can be approximated as
\begin{equation}
\langle\tau_{\rm RD}\rangle\approx \tau_{\rm RD}^0\,
\left(M_{\rm H}\over 10^{13}\, M_{\odot}\right)^{2/3}\,
\left(1+z\over 4\right),
\end{equation}
implying $1-{\rm e}^{-\langle\tau_{\rm RD}\rangle}\approx 1$
for massive galaxies and $1-{\rm e}^{-\langle\tau_{\rm
RD}\rangle}\approx \langle\tau_{\rm RD}\rangle\propto M_{\rm
H}^{2/3}$ for less massive galaxies. As expected, most of the
mass flows from the reservoir to the central BH in the final
couple of $e-$folding times. At early times the ratio of the BH
to the stellar mass is predicted to be much lower than the
final value (cf. Fig.~A1).

In Fig.~A1 we show, as illustrative examples, the evolutions
with galaxy age of the stellar, reservoir and BH masses, and of
the star formation rate (SFR) for galaxies with halo masses of
$10^{13}\, M_{\odot}$ and $10^{11}\,M_{\odot}$, formed at
$z_{\rm form}=3$.

\end{document}